\documentclass[aps,prd,groupedaddress]{revtex4} 
\usepackage{epsfig}
\usepackage{amsfonts}
\usepackage{amssymb}
\usepackage{amsmath}
\usepackage{bm}
\usepackage{latexsym}
\usepackage{mathbbol}
\usepackage{graphicx}
\newcommand{\ud}{\mathrm{d}}
\newcommand{\kslash}{k\!\!\!/}
\newcommand{\pslash}{p\!\!\!/}
\newcommand{\im}{\textrm{Im}}
\begin{document}
\title{The box diagram in Yukawa theory}
\author{ Bernard L. G. Bakker$^{a}$, Jorn K. Boomsma$^{a}$ and Chueng-Ryong Ji$^{b}$\\
$^a$ Department of Physics and Astrophysics, Vrije Universiteit, \\
     De Boelelaan 1081, NL-1081 HV Amsterdam, \\
     The Netherlands\\
$^b$ Department of Physics, \\
North Carolina State University,\\
Raleigh, NC 27695-8202, \\
USA}
\begin{abstract}
We present a light-front calculation of the box diagram in Yukawa
theory.  The covariant box diagram is finite for the case of spin-1/2
constituents exchanging spin-0 particles. In light-front dynamics, 
however, individual time-ordered diagrams are divergent. We analyze
the corresponding light-front singularities and show the equivalence
between the light-front and covariant results by taming the singularities.  
\end{abstract}

\maketitle

\section{Introduction\label{sect.I}}

One of the important theoretical tools to study relativistic
bound-state problems is the light-front quantization method which
provides one of the three forms~\cite{Dirac} of Hamiltonian dynamics
and a Fock-state representation at equal light-front time $\tau = t +
z/c$~\cite{BPP}.  Distinguished features of light-front
dynamics (LFD) may be summarized in two fundamental areas of physics,
namely the vacuum and symmetry in Minkowski space.  Except for zero-modes,
the vacuum of LFD is much simpler than the vacuum of instant form
dynamics (IFD) based on the ordinary equal $t$ quantization.  The
rational energy-momentum dispersion relation in LFD may be the key to
understand this highly non-trivial property of the vacuum. The number
of kinematic generators for the Poincar\'{e} symmetry is also maximal in
LFD and the conversion of dynamical generators from the boost to the
transverse rotation occurs upon the replacement of IFD by
LFD~\cite{JM}.  The compact nature of the rotation may be regarded as an
%advantageous property of the LFD in handling complicate dynamical
advantageous property of the LFD in handling complicated dynamical
operators~\cite{JKM}.

The momentum-space bound-state solutions to a system of relativistic
equations $\psi (x, {\vec k}_\perp, \lambda)$ are functions of the
light-front variables $x_i = (k^0_i + k_i^z)/(p^0 + p^z)$, ${\vec
k}_{\perp i}$, and the particle helicities $\lambda_i$. They are
suitable for the calculation of physical observables such as hadron
form factors and structure functions.  The first step in solving the
full set of coupled Fock-state equations on the light-front (LF) is to
find a simple analytically tractable equation for the valence,
lowest-particle-number sector, and to develop a systematic scheme of
obtaining the contributions from the higher-particle-number sectors
with a desired accuracy.  A frequently used technique to achieve at
least the first step is the projection of the manifestly covariant
Bethe-Salpeter equation on the hypersurface of LFD~\cite{BJS}.  For
instance, the the LF ladder approximation can be obtained by projecting
the ladder approximation of the Bethe-Salpeter equation. The ladder
approximation to the LF two-body bound-state equation involves only up to
three-body Fock-states and the higher-particle-number sectors
beyond the three-body Fock-states cannot be generated in this
approximation. However, they can be generated if the projection
procedure is applied after the covariant Bethe-Salpeter equation is
iterated once.  As an example, the four-body Fock-state of the
stretched box diagram in LFD can be generated by projecting the
covariant box diagram after iterating once the ladder kernel of the
Bethe-Salpeter equation.  In this sense, the procedures of iteration
and projection do not commute.

A three-dimensional reduction of the two-particle Bethe-Salpeter
equation has been proposed~\cite{SFCS2000} and the reduction of the
two-fermion Bethe-Salpeter equation in the framework of LFD has been
studied for the 3+1 dimensional Yukawa model~\cite{SFCS2001}. We study
here a generalization of this model for which the interaction
Lagrangian density is given by
% 010
\begin{equation}
  L = g_1 {\bar\Psi}_1 \Psi_1 \phi_1 + g_2 {\bar\Psi}_2 \Psi_2 \phi_2,
  + g_{12} ( {\bar\Psi}_1 \Psi_2 \sigma + {\rm h.c.}),
\label{eq.010}
\end{equation}
where the fermions correspond to the fields $\Psi_{1,2}$ with rest
masses $M,\;m$ and the exchanged bosons to the fields $\phi_1$,
$\phi_2$, and $\sigma$ with masses $\mu_1$, $\mu_2$, and $\mu$,
respectively.  A physical system where this interaction would be
applicable is the coupled $NN - N^* N^*$ system.

The LF treatment yields three-dimensional quantities for the transition
matrix and the bound-state wave function.  
Since the kernel of the LF Tamm-Dancoff~\cite{PHW} reduced
Bethe-Salpeter equation for the vertex function for the
one-boson-exchange interaction in ladder approximation has a divergence
problem, the introduction of a counterterm to renormalize the integral
equation was proposed~\cite{GHPSW}. In Ref.~\cite{SFCS2001}, the
authors were concerned with the origin of the perturbative counterterms
of the LF ladder Bethe-Salpeter equation for the Yukawa model and
showed that the kernel of the auxiliary integral equation, expanded up
to the fourth power of the coupling constant, $g^4$, naturally yielded
the box counterterm~\cite{GHPSW} and a well defined finite part. As the
authors explained, this is because the perturbative expansion of the LF
scattering amplitude in powers of the coupling constant, obtained from
%the LF T-matrix equation with the kernel calculated up to the same
the LF $T$-matrix equation with the kernel calculated up to the same
order, necessarily reproduces the perturbative covariant ladder
scattering amplitude at that order in $g$. An important necessary
inclusion, in addition to the instantaneous contributions, is the
higher Fock components by going beyond the Tamm-Dancoff approximation.
For example, in Ref.~\cite{SFCS2001}, the divergence could not be
removed if the four-body Fock-component represented by the simultaneous
propagation of two $\sigma$s and two fermions occuring between 
the creation and annihilation of the bosons was not included. Even
higher Fock-state contributions such as the simultaneous propagation of
three $\sigma$s and two fermions should be necessary if terms of order
$g^6$ are included in the kernel. Such a requirement of including
higher Fock-state contributions to remove the relevant divergence in
LFD would persist if higher and higher Fock sectors would be included.
Thus, it is not clear how the reduction program would work in practice
even if the expansion in powers of the coupling constant allows a
%defined number of boson exchanges.
definit number of boson exchanges.

Moreover, LF perturbation theory shows singularities which have nothing
%to do with truncation of the Fock space. Many pitfalls and treacherous
to do with the truncation of Fock space. Many pitfalls and treacherous
points exist in LFD, which is full of surprises. In particular, we
recently discussed the arc contribution and the point singularity in
the contour integration of the LF energy variable~\cite{BDJM}.  Also,
we showed an explicit example of the anomaly associated with the
quantum field theoretic infinities~\cite{BJ2005} that make the
prediction of physical quantities different between the manifestly
covariant approach and LFD even after the amplitude is renormalized,
unless the anomaly-free condition is imposed. Our vector anomaly
analysis~\cite{BJ2005} provided a bottom-up fitness test of the
Standard Model and a model-independent proof of the zero-mode
contribution even in the good (or plus) current matrix element
(helicity zero-to-zero amplitude). Thus, it is significant to further
analyze the LFD singularities.

In this work, we use the generalized Yukawa model given by
Eq.~(\ref{eq.010}) as a testbed for different ways to remove LF
singularities and discuss another type of singularity which is
significant to the bound state problem in LFD.  In LFD one would like
to use the LF time-ordered one-boson-exchange amplitudes as driving
terms in the bound-state equation, similarly to the covariant
one-boson-exchange amplitudes which play the same role in the ladder
Bethe-Salpeter equation.  Here a difficulty arises that can be simply
formulated in perturbation theory:  The manifestly covariant box
diagram in the Yukawa model is finite, whereas the corresponding LF
diagrams are divergent. In this type of diagrams, the residue
calculus is correct and no arc contributions occur in the LF energy
contour integration. Moreover, the LF divergences
cancel~\cite{MvI2005} since this type of singularities corresponds to
finite integrals over the LF energy variable. However, in order to
learn how to exploit the property of cancelling divergences, we
analyzed in detail the box diagram.  We present both the manifestly
covariant calculation and the LF calculation to verify the equivalence
between the two.  Using the integrals over the Feynman parameters in
the manifestly covariant calculation, we expand the on-shell scattering
amplitude in terms of form factors that are functions of the Mandelstam
variables.

In order to be able to calculate the LF amplitudes, we need to
introduce regularization. Two methods were used, namely Pauli-Villars
(PV) regularization where one PV boson was introduced, and dimensional
regularization in the transverse-momentum integrals (DR${}_2$). We
compare these two methods, because we have seen before~\cite{BJ2005}
that in some cases the finite parts of the regulated amplitudes may
differ.  Our LF calculation reveals the zero-mode in the stretched box
depending on the kinematics. We discuss a few different prescriptions
to compute the form factors. We have found that they are prescription
independent.

In Section II, we briefly review the scalar box diagram and introduce
the variables that are used in the rest of the paper. In Section III,
our manifestly covariant calculation is presented for the scattering of
two spin-1/2 particles.  The relation of the matrix elements in spin
space to the invariant form factors is given here too.  We show how,
for a given kinematics, the form factors can be extracted from the
matrix elements.  In Section IV, our LF calculation is presented and
the zero-mode contribution from the stretched box is also discussed.
In Section V, we sketch the two regularization methods we use in the LF
case.  Section VI contains a discussion of the numerical results and our
Conclusions follow in Section VII. In the Appendices A and B, details of
the helicity spinors and the integrals used in DR${}_2$ are given,
respectively.

\section{Analysis of the scalar diagram\label{sect.II}}

\begin{figure}
\begin{center}
 \epsfig{figure=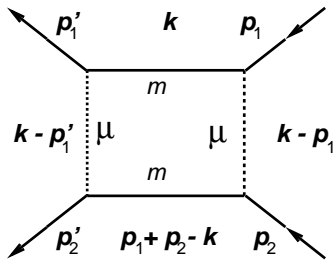,width=50mm}
\caption{The covariant box diagram with momenta defined. 
\label{fig.1}}
\end{center}
\end{figure}

In Refs.~\cite{KSW1959} and \cite{ELOP1966}, a general framework is
given for the calculation of the scalar box diagram. Although these
authors were chiefly interested in the analytic properties of
amplitudes in strong-interaction theory, we may use their methods to
determine the values of the Mandelstam variables for which the box
diagram is non-singular. 

The scalar box diagram is defined by the integral
% 020
\begin{equation}
 I = \int \ud^4 k \int d\alpha_1 \dots d\alpha_4 
 \frac{\delta(1-\alpha_1 \dots -\alpha_4)}
 {[\alpha_1(q^2_1 - m^2_1) + \dots + \alpha_4(q^2_4 - m^2_4)]^4}.
\label{eq.020}
\end{equation}
Of course, Feynman parametrization was used here. 
The momenta $q_i$ on the internal lines are (see Fig.~\ref{fig.1})
% 030
\begin{equation}
 q_1 = k - p_1,\; q_2 = k,\; q_3 = k- p'_1,\; q_4 = p_1 + p_2 - k.
\label{eq.030}
\end{equation}
The corresponding masses are
% 040
\begin{equation}
 m_1 = m_3 = \mu,\; m_2 = m_4 = m.
\label{eq.040}
\end{equation}
The function in the
denominator can be written as $k^{\prime\, 2} + M^2_{\rm cov}$ if the shift
% 060
\begin{equation}
 k \to k' = k - (\alpha_1 + \alpha_4) p_1 - \alpha_4 p_2 - \alpha_3 p'_1
\label{eq.060}
\end{equation}
is performed. 

Next, we make the restriction that the external momenta are on shell for
external particles with mass $M$.  Owing to four-momentum
conservation we can express the momentum dependence in terms of the
three Mandelstam variables $s$, $t$, and $u$, defined as usual:
% 080
\begin{eqnarray}
 s & = & (p_1 + p_2)^2 = (p'_1 + p'_2)^2, 
\nonumber \\
 t & = & (p'_1 - p_1)^2 = (p'_2 - p_2)^2,
\nonumber \\
 u & = & (p_1 - p'_2)^2 = (p'_1 - p_2)^2.
\label{eq.080}
\end{eqnarray}
Upon substitution of the on-shell relations for the external momenta
and the Mandelstam variables $s$ and $t$ we find
% 100
\begin{eqnarray}
 M^2_{\rm cov} & = & (\alpha_2 + \alpha_4) m^2 + (\alpha_1 + \alpha_3) \mu^2
\nonumber \\
 & & -(\alpha_1 + \alpha_3)(\alpha_2 + \alpha_4) M^2 
 -\alpha_2 \alpha_4 s - \alpha_1 \alpha_3 t.
\label{eq.100}
\end{eqnarray}
We consider in this work a situation that is similar to the bound-state
case in the sense that the corresponding scattering amplitude is real,
namely we consider masses $m$ and $M$, related to the fermions in the
loop and the external fermions respectively, such that $M < m$, which
gives us a window $4 M^2 \leq s \leq 4 m^2$ where the contribution from
the $\Psi_2\Psi_2$-intermediate state to the scattering amplitude is
real. While working in this window, we can avoid the complications
caused by the unitarity cuts, when we calculate the diagram for the
process $\Psi_1\Psi_1 \to \Psi_2\Psi_2 \to \Psi_1\Psi_1$.  In that case
the analysis of Refs.~\cite{KSW1959,ELOP1966} shows that the minimum
value of the denominator function defined above is greater than zero in
the domain of integration.

\section{Covariant calculation of the Yukawa box\label{sect.III}}

In this section we discuss the general formalism for the calculation of
the box diagram. First, we sketch the calculation of the amplitude in
spin space and next we connect these matrix elements to invariant form
factors.

\subsection{Amplitude}
\label{Sect.III.A}

We consider the scattering of two spin-1/2 paricles with mass $m$, that
exchange  scalar particles of mass $\mu$. The amplitude can be written as
% 110
\begin{equation}
 {\cal T}_{fi} =
 \bar{u}(p'_1, s'_1)\bar{u}(p'_2, s'_2) {\cal M} u(p_1,s_1) u(p_2,s_2),
\label{eq.110}
\end{equation}
where ${\cal M}$ is a matrix in spin space and depends on the
invariants that can be built from the momenta $p_1$, $p_2$, $p'_1$, and
$p'_2$.

The spin matrix ${\cal M}$, defined in Eq.~(\ref{eq.110}), can be
obtained using the standard Feynman rules. Then one finds
% 120
\begin{eqnarray}
 \mathcal{M}  & = & \int \frac{\ud^4 k}{(2 \pi)^4}
 \frac{(\kslash + m)\otimes(\pslash_1 + \pslash_2 - \kslash + m)}
 {(k^2 - m^2 + i \epsilon)((k - p_1)^2 - \mu^2 + i \epsilon)}
\\ 
\nonumber
 & & \times 
 \frac{1}{((p_1+p_2 - k)^2 - m^2 + i \epsilon) ((k-p'_1)-\mu^2 + i \epsilon)},
\label{eq.120}
\end{eqnarray}
where the notation $\otimes$ is introduced to distinguish the internal
lines connecting $p_i$ and $p'_i$ for $i=$ 1 and 2.
This expression can be rewritten using Feynman parameters as
% 130
\begin{equation}
 \mathcal{M}  = 6 \int_T \ud \alpha_1 \cdots \ud \alpha_4
 \int \frac{\ud^4 k}{(2 \pi)^4} \,
 \frac{(\kslash + m)\otimes(\pslash_1 + \pslash_2 - \kslash + m)}{D^4},
\label{eq.130}
\end{equation}
where
% 140
\begin{eqnarray}
 D & = & \alpha_1 ((k-p_1)^2 - \mu^2) ) + \alpha_2 (k^2 - m^2)\
\nonumber \\
 & & + \alpha_3 ((k - p'_1)^2 - \mu^2 ) + \alpha_4  ((p_1 + p_2 - k)^2-m^2)
 + i \epsilon.
\label{eq.140}
\end{eqnarray}
The domain of integration $T$ is the interior of a regular tetrahedron
given by $T = \{\alpha_i \ge 0, \alpha_1 + \alpha_2 + \alpha_3 +
\alpha_4 = 1\}$.
 
After a Wick rotation it is clear that the amplitude is given by a
convergent integral. Consequently, we are allowed to shift the
integration variable. The shift is of course the same as in the scalar
%case, Eq.~(\ref{eq.060}).  After the shift that produces a denominator
case, Eq.~(\ref{eq.060}).  After the shift, which produces a denominator
that is an even function of the integration variable $k'$, we may
invoke symmetry to prove that the terms in the numerator proportional
to $k'$ vanish upon integration and that the term $k'_\mu k'_\nu$ is
equivalent to a term $g_{\mu\nu} \, k^{\prime 2}/4$.

%The numerator is changed to (we drop the terms linear in $k'$ for symmetry reason)
The numerator is changed to (we drop the terms linear in $k'$)
% 150
\begin{equation}
 - \gamma(1) \cdot k' \; \gamma(2) \cdot k'
 + \left\{\gamma(1) \cdot
 [(1-\alpha_3 -\alpha_4) p_1 + \alpha_2 p_2 + \alpha_3 p'_1] +m \right\}
 \;\left\{\gamma(2) \cdot [(\alpha_3 + \alpha_4)p_1 +
 (1-\alpha_2) p_2 - \alpha_3 p'_1] +m \right\} .
\label{eq.150}
\end{equation}
The gamma-matrices are associated with two different particles: $\gamma(i)$
is associated with the internal line connecting $p_i$ and $p'_i$.

We need the following integrals
% 170
\begin{eqnarray}
 D_0 = 6 \int_T d\alpha \int \frac{\ud^4 k}{(2\pi)^4}
 \frac{1}{(k^2 - M^2_{\rm cov})^4},
\nonumber \\
 D_{\alpha_i} = 6 \int_T d\alpha \int \frac{\ud^4 k}{(2\pi)^4}
 \frac{\alpha_i}{(k^2 - M^2_{\rm cov})^4},
\nonumber \\
 D_{\alpha_i\alpha_j} = 6 \int_T d\alpha \int \frac{\ud^4 k}{(2\pi)^4}
 \frac{\alpha_i\alpha_j}{(k^2 - M^2_{\rm cov})^4},
\nonumber \\
 D_2 = 6 \int_T d\alpha \int \frac{\ud^4 k}{(2\pi)^4}
 \frac{k^2}{(k^2 - M^2_{\rm cov})^4}.
\label{eq.170}
\end{eqnarray}
Upon performing the integral over the momentum and using a Wick
rotation, we find
% 200
\begin{eqnarray}
 D_{\alpha_i} = \frac{i}{(4\pi)^2} \int_T d\alpha
 \frac{\alpha_i}{M^4_{\rm cov}},
\nonumber \\
 D_{\alpha_i\alpha_j} = \frac{i}{(4\pi)^2} \int_T d\alpha
 \frac{\alpha_i\alpha_j}{M^4_{\rm cov}},
\nonumber \\
 D_2 = \frac{-2i}{(4\pi)^2} \int_T d\alpha \frac{1}{M^2_{\rm cov}}.
\label{eq.200}
\end{eqnarray}
Owing to the symmetries of the diagram the mass function is symmetric
under the transpositions $\alpha_1 \leftrightarrow \alpha_3$ and
$\alpha_2 \leftrightarrow \alpha_4$.
Using this symmetry of the denominator, we find the identities
% 180
\begin{eqnarray}
 D_0 & = & 2 (D_{\alpha_1} + D_{\alpha_2}),
\nonumber \\
 D_{\alpha_1} & = & D_{\alpha_3},\quad D_{\alpha_2} = D_{\alpha_4},
\nonumber \\
 D_{\alpha^2_1} & = & D_{\alpha^2_3},\quad D_{\alpha^2_2} = D_{\alpha^2_4},
\nonumber \\
 D_{\alpha_1 \alpha_2} & = & D_{\alpha_1 \alpha_4} = D_{\alpha_2 \alpha_3} = 
 D_{\alpha_3 \alpha4}.
\label{eq.180}
\end{eqnarray}
Using the fact that the alpha's add up to 1, we can derive two more
relations, namely
% 190
\begin{eqnarray}
 D_{\alpha^2_1} & = & D_{\alpha_1} - 2 D_{\alpha_1 \alpha_2} - D_{\alpha_1 \alpha_3},
\nonumber \\
 D_{\alpha^2_2} & = & D_{\alpha_2} - 2 D_{\alpha_1 \alpha_2} - D_{\alpha_2 \alpha_4}.
\label{eq.190}
\end{eqnarray}
Using the symmetries just discussed, we find that only $D_{\alpha_1}$,
$D_{\alpha_2}$, $D_{\alpha_1\alpha_2}$, $D_{\alpha_1 \alpha_3}$,
$D_{\alpha_2 \alpha_4}$, and $D_2$ are independent.

\subsection{Spin Structure}
\label{Sect.III.2}

We now gather the different pieces of the amplitude. We write
% 210
\begin{equation}
 {\cal M} = - \gamma(1) \cdot \gamma(2) \,\mbox{\small $\frac{1}{4}$} D_2 
 + m\, \gamma(1) \cdot D(1) + m\, \gamma(2) \cdot D(2) +
 \gamma(1)_\mu \gamma(2)_\nu\, D(12)^{\mu\nu} + m^2\, D_0,
\label{eq.210}
\end{equation}
where we used the vectors
% 220
\begin{eqnarray}
 D(1) & = & \mbox{\small $\frac{1}{2}$} \left[ P D_0
 + q D_{\alpha_1}
 - q'D_{\alpha_1} \right],
\nonumber \\
 D(2) & = & \mbox{\small $\frac{1}{2}$} \left[ P D_0
 - q D_{\alpha_1}
 + q'D_{\alpha_1} \right],
\nonumber \\
 & = & P D_0 - D(1),
\label{eq.220}
\end{eqnarray}
and the tensor
% 230
\begin{eqnarray}
 D(12)^{\mu\nu} & = & \; \frac{1}{2} P^\mu P^\nu \left[ D_{\alpha_1} 
 + 2 D_{\alpha_1 \alpha_2} + 2 D_{\alpha_2 \alpha_4} \right] 
\nonumber \\
% & & + \frac{1}{4} q^\mu q^\nu \left[ -D_{\alpha_1} 
 & & + \frac{1}{4} \left[ q^\mu q^\nu + q'^\mu q'^\nu \right]
 \left[ -D_{\alpha_1} + 2 D_{\alpha_1 \alpha_2} + D_{\alpha_1 \alpha_3} \right] 
%\nonumber \\
% & & + \frac{1}{4} q'^\mu q'^\nu \left[ -D_{\alpha_1} 
% + 2 D_{\alpha_1 \alpha_2} + D_{\alpha_1 \alpha_3} \right]
\nonumber \\
 & & - \frac{1}{4} P^{[\mu} q^{\nu]} D_{\alpha_1}
 - \frac{1}{4} P^{[\mu} q'^{\nu]}  D_{\alpha_1}
 - \frac{1}{4} q^{\{\mu} q^{\prime\,\nu\}}
 D_{\alpha_1 \alpha_3}.
%\nonumber \\
% & & - \frac{1}{4} q^{\{\mu} q^{\prime\,\nu\}}
% D_{\alpha_1 \alpha_3}.
\label{eq.230}
\end{eqnarray}
Here we have used the vectors $P = p_1 + p_2$, $q = p_1 - p_2$ and $q' = p'_1 - p'_2$ to get symmetric expressions. We use the notation $p^{\{\mu} q^{\nu\}}= p^\mu q^\nu + q^\mu p^\nu$ and $p^{[\mu} q^{\nu]}= p^\mu q^\nu - q^\mu p^\nu$.

In view of the symmetries of the spin matrix elements of the identity
and the $\gamma$-matrices (see Appendix \ref{sect.A}), we find that
the matrix ${\cal T}$ has the following structure
% 240
\begin{equation}
 {\cal T} =
 \left(
 \begin{array}{rrrr}
 {\cal T}_{11} & {\cal T}_{12} & {\cal T}_{13}  & {\cal T}_{14}  \\
 {\cal T}_{21} & {\cal T}_{22} & {\cal T}_{23}  & {\cal T}_{24}  \\
 -{\cal T}^*_{24} & {\cal T}^*_{23} & {\cal T}^*_{22}  & -{\cal T}^*_{21}  \\
 {\cal T}^*_{14} & -{\cal T}^*_{13} & -{\cal T}^*_{12}  & {\cal T}^*_{11}
 \end{array}
 \right).
\label{eq.240}
\end{equation}
Here we use the following numbering of the two-fermion spin states
% 250
\begin{equation}
 |1\rangle = |\uparrow\uparrow\rangle,
 |2\rangle = |\uparrow\downarrow\rangle,
 |3\rangle = |\downarrow\uparrow\rangle,
 |4\rangle = |\downarrow\downarrow\rangle.
\label{eq.250}
\end{equation}
Clearly, there are eight independent complex matrix elements which
correspond to sixteen independent real numbers. 

Upon taking matrix elements between spinors and using the Dirac
equation to simplify some matrix elements, we find the following structure
% 260
\begin{eqnarray}
 {\cal T} & = & O^1 F_1 + O^2 F_2 + O^3 F_3 + O^4 F_4,
\nonumber \\
 O^1 & = & I(1) \otimes I(2), 
 \hspace{12.7em} F_1 = 2 m M \left( 2 D_{\alpha_1} + D_{\alpha_2} \right) + M^2 \left( 2 D_{\alpha_1} + D_{\alpha_2 \alpha_4} \right) + 2 m^2 \left(D_{\alpha_1} + D_{\alpha_2} \right),
\nonumber \\
 O^2 & = & 
 [ I(1) \otimes (p_1 \cdot \Gamma(2)) + (p_2 \cdot \Gamma(1)) \otimes I(2) ] ,
 \hspace{1em} F_2 = m D_{\alpha_2} + M \left( 2 D_{\alpha_1 \alpha_2} + D_{\alpha_2 \alpha_4} \right)  ,
\nonumber \\
 O^3 & = & (p_2 \cdot \Gamma(1)) \otimes (p_1 \cdot \Gamma(2)), 
 \hspace{7.6em} F_3 = D_{\alpha_2 \alpha_4},
\nonumber \\
 O^4 & = & \Gamma(1) \cdot \Gamma(2),
 \hspace{13.0em} F_4 = -D_2/4.
\label{eq.260}
\end{eqnarray}
Apparently, there are four independent form factors. The reason for this
small number is that there is a high degree of symmetry in the
mass function $M^2_{\rm cov}$. 	If this symmetry would be broken by
e.g. differences in masses of the particles, more form factors would occur.
Remarkably, the integral $D_{\alpha_1 \alpha_3}$ does not occur.

Using this representation, one can easily derive linear relations
between the spin matrix elements if the matrix elements of the spin
operators $O^1 \dots O^4$ are known.

\subsection{Extraction of form factors\label{sect.III.C}}

In a manifestly covariant calculation, one does not need the
spin-matrix elements ${\cal T}_{fi}$ to extract the form factors as
Eq.~(\ref{eq.260}) shows. However, in a LF calculation, which breaks
manifest covariance, there does not exist a relation like
Eq.~(\ref{eq.260}). Thus, one must try to solve the linear relation 
between
the ${\cal T}_{fi}$ and the $F_j$ in a specific kinematics.  We work in
the center of mass system (CMS). Of course, other kinematics will also enable
%us to extract the form factors, but the CMS that is detailed below
us to extract the form factors, but the CMS, which is detailed below,
seems to be the simplest one.

For any kinematics, we may write
% 270
\begin{equation}
 {\cal T}_{fi} = \sum_j O^j_{fi}\,F_j.
\label{eq.270}
\end{equation}
The extraction procedure consists in choosing a set of spin labels
$\{f_l,i_l\}, \; l=1,\dots, 4$, such that the square matrix
$O^j_{f_l,i_l}, \; j=1,\dots 4,\; l=1, \dots, 4$, is nonsingular. An
obvious choice is fixing the row label $f$ to 1 or 2 and let the column
index $i$ run from 1 to 4. This choice was made in this work and we
have found that it gives unambiguous results as long as the sine of the
scattering angle $\theta$ and the three-momentum $p$ of the scattered
particles do not vanish.

As an illustration of the problems one may encounter otherwise, we show
the structure of the amplitude matrix in forward ($\theta=0$) and
backward ($\theta = \pi$) kinematics. In forward kinematics, the
amplitude $\mathcal{T}$ reduces to a diagonal form
% 280
\begin{equation}
 {\mathcal{T}}(\theta = 0) =
 \left(
 \begin{array}{rrrr}
 {\cal T}_{11} & 0 & 0 & 0  \\
 0 & {\cal T}_{11} & 0 & 0  \\
 0 & 0 & {\cal T}_{11} & 0  \\
 0 & 0 & 0 & {\cal T}_{11}
 \end{array}
 \right).
\label{eq.280}
\end{equation}
Clearly, there is only one independent amplitude in forward kinematics
and one obviously cannot extract four form factors at $\theta=0$.

In backward kinematics, the amplitude $\mathcal{T}$ is a slightly more
complicated and has the following structure
% 290
\begin{equation}
 {\mathcal{T}}(\theta = \pi) =
 \left(
 \begin{array}{rrrr}
 {\cal T}_{11} & 0 & 0 & 0  \\
 0 & {\cal T}_{11} & {\cal T}_{23} & 0  \\
 0 & {\cal T}_{23} & {\cal T}_{11} & 0  \\
 0 & 0 & 0 & {\cal T}_{11}
 \end{array}
 \right).
\label{eq.290}
\end{equation}
Now two independent amplitudes occur, which also does not allow to
extract the four form factors.

In the LF case, we may use the same matrices $O^j_{fi}$ to connect the LF
amplitude ${\cal T}^d_{fi}$ corresponding to diagram $d$ to an LF `form
factor' $F^d_j$. We use here the terminology of form factors, although
the quantities $F^d_j$ are not invariant objects. We shall, however,
show that upon regularization of the LF amplitudes, the corresponding
LF form factors add up to the invariant ones, viz.
% 300
\begin{equation}
 \sum_d F^d_j = F_j.
\label{eq.300}
\end{equation}

\subsection{Kinematics}

In the box diagram that depends on three independent momenta, a
simplification cannot be achieved by choosing e.g. the Breit frame,
which appeared to be so helpful in, for instance, the triangle diagram.
Therefore, we choose just the CMS. A slight simplification can be
achieved if one limits the incoming and outgoing momenta to the
$xz$-plane. Here we follow Erkelenz~\cite{E1974}. In particular, we choose
% 310
\begin{equation}
 {\bm p}_1 = p \hat{\bm e}_z, \;
 {\bm p}^{\;\prime}_1 =
 p ( \sin\theta \cos\phi, \sin\theta \sin\phi , \cos\theta), \quad
 {\bm p}_2 = - {\bm p}_1,\; {\bm p}^{\;\prime}_2 = - {\bm p}^{\;\prime}_1.
\label{eq.310}
\end{equation}
In order to construct the correct helicity spinors in initial and final 
states,
it is important to define the polar and azimuthal angles correctly. They are
% 320
\begin{equation}
 {\bm p}_1:\; \theta_1 = 0, \phi_1 = 0;\;
 {\bm p}_2:\;\theta_2 = \pi, \phi_2 = \pi, \quad
 {\bm p}'_1:\; \theta'_1 = \theta, \phi'_1 = 0;\;
 {\bm p}_2:\;\theta'_2 = \pi-\theta, \phi'_2 = \pi.
\label{eq.320}
\end{equation}
The corresponding spinors can be found in Appendix~\ref{sect.A}.

Here, we want to point out that the zero mode we mentioned in the
Introduction is found in this kinematics when $p^+_1 = p^{\prime +}_1$,
which occurs for $p=0$ or $\theta =0$. If we had chosen instead of the
$xz$-plane the $xy$-plane as the scattering plane, the zero mode would
have occurred for any values of the scattering angle and the
three-momentum.

\section{Light-front calculation\label{sect.IV}}

The invariant amplitude Eq.~(\ref{eq.120}) can be rewritten in LF coordinates
% 340
\begin{equation}
 \mathcal{M} = \int \frac{\ud^2 k_\perp}{(2\pi)^2}\int \frac{\ud k^+}{2 \pi}
 \int \frac{\ud k^-}{2 \pi} \frac{1}{\Phi}
 \frac{(\kslash + m)\otimes(-\kslash + \pslash_1 + \pslash_2 + m)}
 {(k^- - H_1)(k^- - H_2)(k^- - H_3)(k^-  - H_4)},
\label{eq.340}
\end{equation}
where the phase-space factor $\Phi$ is given by
% 350
\begin{equation}
 \Phi = 16 k^+ (k^+ - p_1^+)(k^+ - {p'_1}^+)(k^+ - p_1^+ - p_2^+).
\label{eq.350}
\end{equation}
The `Hamiltonians' $H_i$ are defined as
% 360
\begin{eqnarray}
 H_1 & = & p_1^- +
 \frac{(\vec k^\perp - \vec p^\perp_1)^2 + \mu^2 - i \epsilon}{2 (k^+ - p_1^+)} 
\nonumber \\
 H_2 & = & \frac{(\vec k^\perp)^2 + m^2 - i \epsilon}{2 k^+} 
\nonumber \\
 H_3 & = & {p'_1}^- + \frac{(\vec k^\perp - {\vec p}^{\, \prime \perp})^2
 + \mu^2 - i \epsilon}{2 (k^+ - {p'_1}^+)} 
\nonumber \\
 H_4 & = & p_1^- + p_2^- +
 \frac{(\vec k^\perp - \vec p^\perp_1 - \vec p^\perp_2)^2 + m^2 - i \epsilon}
 {2 (k^+ - p_1^+ - p_2^+)}.
\label{eq.360}
\end{eqnarray}

We choose $p_1^+ \geq {p'_1}^+$ for definitness. In the opposite case,
some details of the calculations would change, but the general picture,
and in particular our conclusions, would remain the same.

In order to get the light-front time-ordered (LFTO) diagrams, we
integrate over $k^-$ first.  The positions of the poles define five
regions in $k^+$, each of them with a different number of poles in the
upper and lower plane. They are
\begin{enumerate}
\item $k^+ < 0: \im H_1 >0, \im H_2 > 0, \im H_3 > 0, \im H_4 > 0$
\item $0 < k^+ < {p'_1}^+: \im H_1 > 0, \im H_2 < 0, \im H_3 > 0, \im H_4 > 0$
\item ${p'_1}^+ < k^+ < p_1^+: \im H_1 > 0, \im H_2 < 0, \im H_3 < 0, \im H_4 >
0$
\item $p_1^+ < k^+ < p_1^+ + p_2^+: \im H_1 < 0, \im H_2 < 0, \im H_3 < 0, \im H
_4 > 0$
\item $k^+ > p_1^+ + p_2^+: \im H_1 < 0, \im H_2 < 0, \im H_3 < 0, \im H_4 < 0$.
\end{enumerate}
In the regions 1 and 5, all poles are in one half of the complex plane
so that the integral over $k^-$ vanishes. In regions 2 and 4, one
of the poles is in one half plane, while the other poles are in the
other half plane. We close the contour in the half plane with one pole
and perform the integration. In region 3, there are two poles in
either half of the complex plane so that for this calculation two
residues have to be included.

\begin{figure}[t]
\begin{center}
\includegraphics[width=80mm]{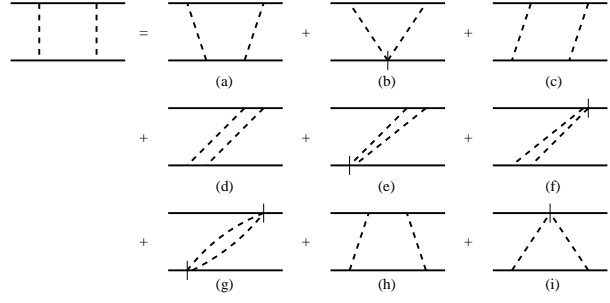}
\end{center}
\vspace{1ex}
\caption{The splitting of the covariant box into light-front time-ordered 
diagrams.  \label{fig.2}}
\end{figure}

Since two of the internal particles are fermions, the instantaneous
parts have to be taken into account.  We obtain the LFTO diagrams of
Fig.~\ref{fig.2} when we perform the described analysis.
Diagrams (a) and (b) correspond to region~2, diagrams (c), (d), (e),
(f), and (g ) correspond to region~3 and diagrams (h) and (i)
correspond to region~4. We can decrease the number of diagrams we have
to use with the blink mechanism~\cite{LB1995}, which we will do for
each region.

\subsection*{Region 2}
\begin{figure}[htb]
\begin{center}
\includegraphics[width=60mm]{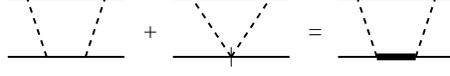}
\end{center}
\caption{The sum of the two diagrams of region 2 is a diagram with a blink.
\label{fig.3}}
\end{figure}

We can combine the propagating diagram with the instantaneous diagram
in region 2 to obtain a diagram with a blink, as shown in
Fig.~\ref{fig.3}. We will call this diagram with the blink diagram 1.

The corresponding amplitude of diagram 1 is
% 370
\begin{equation}
 \mathcal{M}_1 = \int \! \frac{\ud^2 k^\perp}{(2 \pi)^2} \int_0^{{p'}_1^+}
 \!\frac{\ud k^+}{2 \pi} \frac{-i}{\Phi}
 \frac{\left[ \kslash_{\mathrm{on}} + m\right] \otimes
 \left[ \gamma_\mu \left(p_1^\mu + p_2^\mu - k_{\mathrm{on}}^\mu \right)
 + m\right]}{D_1 D_2 D_3},
\label{eq.370}
\end{equation}
where the momenta with subscript `on' are four-momenta with the
minus-component calculated from the on-mass-shell condition, i.e.,
$k^-_{\rm on} = ((\vec{k}^\perp)^2 + m^2)/2 k^+$.  The energy
denominators are
% 380
\begin{eqnarray}
 D_1 & = H_2 - H_1 = & {p'_1}^- - \frac{(\vec k^\perp)^2 + m^2}{2 k^+}
 - \frac{(\vec p^{\, \prime \perp}_1 - \vec k^\perp)^2 + \mu^2}
 {2 \left({p'_1}^+ - k^+\right)} 
\nonumber \\
 D_2 & = H_4 - H_1 = & p_1^- + p_2^- - \frac{(\vec k^\perp)^2 + m^2}{2 k^+}
 - \frac{(\vec p^\perp_1 + \vec p^\perp_2 - \vec k^\perp) + m^2}
 {2(p_1^+ + p_2^+ - k^+)} 
\nonumber \\
 D_3 & = H_3 - H_1 = & p_1^- - \frac{(\vec k^\perp)^2 + m^2}{2 k^+} -
 \frac{(\vec p_1^\perp - \vec k^\perp)^2 + \mu^2}{2 (p_1^+ - k^+)}.
\label{eq.380}
\end{eqnarray}
The phase-space factor is
% 390
\begin{equation}
 \Phi = 16 k^+ (k^+ - p_1^+)(k^+ - {p'_1}^+)(k^+ - p_1^+ - p_2^+).
\label{eq.390}
\end{equation}

\subsection*{Region 3}
In this region two poles contribute, which are not directly coupled to
diagrams, see Ref.~\cite{LB1995} for a discussion. Each of the diagrams is a
linear combination of the corresponding residues. The diagrams
contributing in this region are (c), (d), (e), (f), and (g) in
Fig.~\ref{fig.2}.
\begin{figure}[htb]
\begin{center}
 \includegraphics[width=80mm]{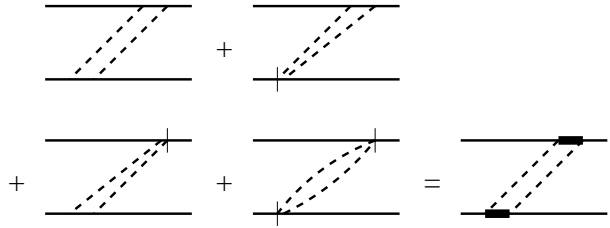}
\end{center}
\caption{The sum of the instantaneous diagrams of region 3 and the
 stretched box is the stretched box with two blinks.
\label{fig.4}}
\end{figure}
The open diamond, diagram (c), does not contain any blinks. We will
call this diagram~2. Its amplitude is given by
% 400
\begin{equation}
 \mathcal{M}_2 = \int \! \frac{\ud^2 k^\perp}{(2 \pi)^2} \int_{{p'}_1^+}^{p_1^+}
 \!\frac{\ud k^+}{2 \pi} \frac{i}{\Phi}
 \frac{\left[\kslash_{\mathrm{on}} + m\right] \otimes
 \left[\gamma_\mu (p_1 + p_2 - k)_{\mathrm{on}}^\mu + m\right]}{D'_1 D_2 D_3}.
\label{eq.400}
\end{equation}
The instantaneous diagrams (e), (f), and (g) can be combined with the
stretched box diagram (d), to create a stretched box with two blinks
(see Fig.~\ref{fig.4}).  The corresponding amplitude is
% 410
\begin{equation}
 \mathcal{M}_3 = \int \!\frac{\ud^2 k^\perp}{(2 \pi)^2}
 \int_{{p'}_1^+}^{p_1^+} \!\frac{\ud k^+}{2 \pi} \frac{i}{\Phi}
 \frac{\left[\gamma_\mu (p_1^\mu-(p_1-k)_{\mathrm{on}}^\mu) + m\right]\otimes
\left[\gamma_\nu ({p'_2}^\nu-(k-p'_1)_{\mathrm{on}}^\nu) + m \right]}
 {D'_1 D'_2 D_3} ,
\label{eq.410}
\end{equation}
where
% 420
\begin{eqnarray}
 D'_1 = H_4 - H_2 & = & {p'_2}^- -
 \frac{(\vec p_1^\perp + \vec p_2^\perp - \vec k^\perp)^2 + m^2}
 {2(p_1^+ + p_2^+ - k^+)}
 - \frac{(\vec k^\perp - \vec p^{\, \prime \perp}_1 )^2 + \mu^2}
 {2 (k^+ - {p'_1}^ +)} 
\nonumber \\
 D_2' = H_3 - H_2 & = & p'^-_2 - p_2^- -
 \frac{(\vec k^\perp - \vec p^{\, \prime \perp}_1 )^2 + \mu^2}{2(k^+ - p'^+_1)}
 - \frac{(\vec p_1^\perp - \vec k^\perp)^2 + \mu^2}{2 (p_1^+ - k^+)}.
\label{eq.420}
\end{eqnarray}
We will call the stretched box with blinks diagram 3. This diagram has a
subtlety, i.e., it contributes even when the integration region goes to 
zero;
this is called a zero mode. In Sec.~\ref{sect.V.1} we will elaborate on
this subject.

\subsection*{Region 4}
\begin{figure}[b]
\begin{center}
 \includegraphics[width=80mm]{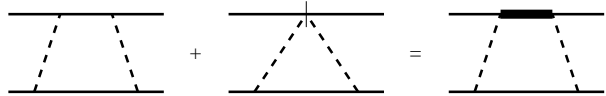}
\end{center}
\caption{The sum of the instantaneous and propagating diagrams in region 4
 is equal to a diagram with a blink.  \label{fig.5}}
\end{figure}

In this region, only the pole corresponding to $H_4$ contributes to the
amplitude so that we now have diagrams (h) and (i) in Fig.~\ref{fig.2}. 
Again we can sum these two diagrams to obtain the diagram with the 
blink (see Fig.~\ref{fig.5}). We will refer to this diagram as diagram 4.
%The amplitude corresponding to diagram 4 is
The corresponding amplitude is
% 430
\begin{equation}
 \mathcal{M}_4 = \int \!\frac{\ud^2 k^\perp}{(2 \pi)^2}
 \int_{p_1^+}^{p_1^+ + p_2^+} \! \frac{\ud k^+}{2 \pi} \frac{-i}{\Phi}
 \frac{\left[\gamma_\mu (p_1^\mu+p_2^\mu-(p_1 + p_2 - k)_{\mathrm{on}}^\mu)
 + m \right]
\otimes \left[ \gamma_\nu (p_1 + p_2  - k)_{\mathrm{on}}^\nu + m\right]}
{D'_1 D_2 D''_3} .
\label{eq.430}
\end{equation}
The energy denominator different from those of regions 2 and 3 is
% 440
\begin{equation}
 D''_3 = H_4 - H_1 = p_2^- -
 \frac{(\vec p_1^\perp + \vec p_2^\perp - \vec k^\perp)^2 + m^2}
{2(p_1^+ + p_2^+ - k^+)}
 - \frac{(\vec k^\perp - \vec p_1^\perp)^2 + \mu^2}{2 (k^+ - p_1^+)}.
\label{eq.440}
\end{equation}

\subsection{Amplitude}
When we take the matrix elements of the four $\mathcal{M}_i$ derived
above, between spinors, we end up with the contribution of the diagram
to the amplitude $\mathcal{T}$.  All four diagrams have the same
structure so that we can parametrize the contribution of each diagram
to the total amplitude $\mathcal{T}$ in the following way ($d$ labeling
the four LF contributions)
% 460
\begin{eqnarray}
 \mathcal{T}^d &  = & \overline{u}(p'_1, s'_1) \overline{u}(p'_2, s'_2)
 \mathcal{M}_d u(p_1,s_1) u(p_2, s_2) 
\nonumber \\
 & = & C(d)^{\mu \nu}_{\Gamma \Gamma} \Gamma(1)_\mu \otimes \Gamma(2)_\nu +
 C(d)^\mu_{\Gamma I} \Gamma(1)_\mu \otimes I(2)
 + C(d)^\nu_{I \Gamma} I(1) \otimes \Gamma(2)_\nu 
\nonumber \\
 & & + C(d)_{II} I(1) \otimes I(2)\label{diag_ampl}.
\label{eq.460}
\end{eqnarray}
The tensor $C(d)^{\mu \nu}_{\Gamma \Gamma}$, the vectors
$C(d)^\mu_{\Gamma I}$ and $C(d)^\nu_{I \Gamma}$ and the constant
$C(d)_{II}$ are of course different for each diagram. Taking for
%example the trapezoid diagram of region 2, we obtain
example diagram 1, we obtain
% 470
\begin{eqnarray}
 C(1)^{\mu \nu}_{\Gamma \Gamma} & = &
 \int \! \frac{\ud^2 k^\perp}{(2 \pi)^2} \int_0^{{p'}_1^+}
 \!\frac{\ud k^+}{2 \pi} \frac{-i}{\Phi} \frac{k^\mu_\mathrm{on}
 \left(p^\nu_1 + p^\nu_2 - k^\nu_\mathrm{on} \right)}{D_1 D_2 D_3}, 
\nonumber \\
 C(1)^\mu_{\Gamma I} & = & \int \!\frac{\ud^2 k^\perp}{(2 \pi)^2}
 \int_0^{{p'}_1^+} \!\frac{\ud k^+}{2 \pi} \frac{-i}{\Phi}
 \frac{k^\mu_\mathrm{on} m }{D_1 D_2 D_3}, 
\nonumber \\
 C(1)^\nu_{I \Gamma} & = & \int \!\frac{\ud^2 k^\perp}{(2 \pi)^2}
 \int_0^{{p'}_1^+} \!\frac{\ud k^+}{2 \pi} \frac{-i}{\Phi}
 \frac{m \left(p^\nu_1 + p^\nu_2 - k^\nu_\mathrm{on} \right)}{D_1 D_2 D_3}, 
\nonumber \\
 C(1)_{I I} & = & \int \!\frac{\ud^2 k^\perp}{(2 \pi)^2} \int_0^{{p'}_1^+}
 \!\frac{\ud k^+}{2 \pi} \frac{-i}{\Phi} \frac{m^2 }{D_1 D_2 D_3}.
\label{eq.470}
\end{eqnarray}
The expressions for the other diagrams are similar.

In principle, we now only have to calculate $C(d)^{\mu \nu}_{\Gamma
\Gamma}$, $C(d)^\mu_{\Gamma I}$, $C(d)^\nu_{I\Gamma}$ and $C(d)_{II}$
for each diagram. Using Eq.~(\ref{diag_ampl}) we obtain the
corresponding $\mathcal{T}^d$ and then use Eq.~(\ref{eq.270}) to
extract the form factors. We end up with the contributions of each
diagram to the form factors. 

\subsection{Rationalization}
Because the numerator and denominator contain terms like 
$1 / (p'^+_1)^2$, it seems there are singularities for the $k^+$
integration. But this is not the case and to show this, we rationalize
the denominators. Let us, for example, take the first diagram and define
% 480
\begin{eqnarray}
 d_1 & = & 2 k^+ \left({p'_1}^+ - k^+\right) D_1, 
\nonumber \\
 d_2 & = & 2 k^+ \left(p_1^+ + p_2^+ - k^+\right) D_2, 
\nonumber \\
 d_3 & = & 2 k^+ \left(p_1^+ - k^+\right) D_3.
\label{eq.480}
\end{eqnarray}
These $d_i$'s are rationalized and we can  proceed
to obtain rationalized denominators for each diagram. We now use
for diagram~1
% 490
\begin{equation}
 \frac{1}{\Phi} \frac{1}{D_1 D_2 D_3} = \frac{(k^+)^2}{2}
 \frac{1}{d_1 d_2d_3}
\label{eq.490}
\end{equation}
to rewrite all the integrals in terms of the $d_i$'s. Then, it becomes
clear that the $k^+$-integration is not singular.

\subsection{Zero modes in the box diagram\label{sect.V.1}}

Domain 3, $p^{\prime +} < k^+ < p^+$, corresponds to two LFTO diagrams;
the open diamond and the stretched box. If we combine the
LF-propagating and instantaneous parts to blinks~\cite{LB1995}, the
longitudinal singularities cancel out.  It is easy to show that the
open diamond is obtained by integrating a function that is finite in
domain 3. Thus, its contribution vanishes if this domain shrinks to a
point. The stretched box, however, contains a zero mode.

To show this, we look at the integrand for the stretched box which is
given by Eqs.~(\ref{eq.380}) and (\ref{eq.390}). Upon rationalization, 
we find that the integrand is proportional to ${\cal F}_3$ given by
% 530
\begin{equation}
 {\cal F}_3 = \frac{1}{2} 
 \frac{(p^+_1 - k^+)(k^+ - p^{\prime +}_1) N_3}{d'_1 d'_2 d_3},
\label{eq.530}
\end{equation}
where $N_3$ is the numerator for ${\cal M}_3$.  The factors $(p^+_1 -
k^+)$ and $(k^+ - p^{\prime +}_1)$ are just the ones needed to cancel
the longitudinal singularities in $N_3$. Indeed we find
% 540
\begin{eqnarray}
 (p^+_1 - k^+)  
\left[\gamma_\mu (p_1^\mu-(p_1-k)_{\mathrm{on}}^\mu) + m\right]
 & = & 
 - \frac{1}{2} [({\vec p}^\perp_1 - {\vec k}^\perp)^2 + \mu^2]
 \gamma^+(1) + (p^+_1 - k^+) \times\;
 {\rm finite}\; {\rm terms},
\nonumber \\
 (k^+ - p^{\prime +}_1) 
 \left[\gamma_\nu ({p'_2}^\nu-(k-p'_1)_{\mathrm{on}}^\nu) + m \right]
 & = & - \frac{1}{2}
 [({\vec k}^\perp - {\vec p}^{\,\prime\,\perp}_1)^2 + \mu^2]
 \gamma^+(2) + (k^+ - p^{\prime +}_1)
 \times\; {\rm finite}\; {\rm terms}.
\label{eq.540}
\end{eqnarray}
Thus, we see that upon taking the limit $p^{\prime +}_1 \to p^+_1$ the
numerator reduces to
% 550
\begin{equation}
 N^{\rm zm} = \frac{1}{8} \gamma^+(1) \gamma^+(2)
 [({\vec p}^\perp_1 - {\vec k}^\perp)^2 + \mu^2]
 [({\vec k}^\perp - {\vec p}^{\,\prime\, \perp}_1)^2 + \mu^2].
\label{eq.550}
\end{equation}
Next we take the same limit in the denominators to find
the final result for the zero mode integrand
% 570
\begin{equation}
 {\cal F}^{\rm zm} =
 - \frac{\gamma^+(1)\gamma^+(2)}{8k^+ (p^+_1 + p^+_2 - k^+)}
 \; \frac{1}
 {(k^+ - p^{\prime +}_1) [({\vec p}^\perp_1 - {\vec k}^\perp)^2 + \mu^2]
 + (p^+_1 - k^+)              )
 [({\vec k}^\perp-{\vec p}^{\,\prime\,\perp})^2 + \mu^2]} .
\label{eq.570}
\end{equation}

We see that the zero mode occurs in the $\gamma^+(1) \gamma^+(2)$
tensor element only. Moreover, the logarithmic divergence is clearly
visible: for $|{\vec k}^\perp| \to \infty$, 
${\cal F}^{\rm zm}$ behaves like $1/{\vec k}^{\,\perp\, 2}$.
Finally, the factors $p^+_1 - k^+$  and
$k^+ - p^{\prime +}_1$ in the denominator are just the
factors that cause the zero mode to survive. The integral over
$k^+$ can be done by making the transformation
% 580
\begin{equation}
 k^+ = p^{\prime +}_1 + x (p^+_1 - p^{\prime +}_1).
\label{eq.580}
\end{equation}
Then the zero-mode amplitude is obtained after taking the limit 
$p^{\prime +}_1 \to p^+_1$. It is
% 590
\begin{equation}
 {\cal M}^{\rm zm} = - \frac{\gamma^+(1) \gamma^+(2)}{8p^+_1 p^+_2}
 \frac{1}{(2\pi)^3} \int^1_0 \ud x \int \ud^2 k^\perp
 \frac{1}
 { x({\vec p}^\perp_1 - {\vec k}^\perp)^2 + 
 (1-x)({\vec k}^\perp-{\vec p}^{\,\prime\,\perp}_1)^2 + \mu^2}.
\label{eq.590}
\end{equation}
\noindent
The divergent integral over the transverse momentum can be regulated
using DR${}_2$ or PV.

\section{Regularization methods}
\label{sect.V}

Now that we have the expressions for the four diagrams, we can see that
there are ultraviolet logarithmic divergences in the $\vec
k^\perp$-integration. These divergences have been explored by Van
Iersel~\cite{MvI2004}. The four diagrams all have a term proportional
to $(\vec k^\perp)^4$ in the numerator and terms proportional to $(\vec
%k^\perp)^6$ in the denominator, making the integral divergent.  Van
k^\perp)^6$ in the denominator, making the integral logarithmically divergent.
Van Iersel has shown that the divergences cancel when one sums all the
diagrams, but the question whether the summation of finite parts gives
the covariant results remained open.  We discuss two possible
regularization schemes, two-dimensional dimensional regularization
(DR${}_2$) and Pauli-Villars (PV) regularization.

\subsection{Dimensional regularization\label{dimreg}}

In the LF case, we only need to regulate the transverse directions, so
that we compute the diagram as an analytic function of the
dimensionality $D$ of these directions. The final quantity should then
have a well-defined limit as $D \to 2$; we call this form of
dimensional regularization two-dimensional dimensional regularization
(DR$_2$).

The divergent parts correspond to terms in the numerator proportional
to $\vec k^4_\perp$. In order to use DR$_2$, we have to rewrite the
denominator in the form $D = ((\vec k^\perp)^2 + (M^\perp)^2)^3$. We
will obtain this form by introducing Feynman-parameters followed by a
shift of the integration variable.

The denominators of all the diagrams have the same structure, $D = d_1
d_2 d_3$. We introduce Feynman parameters to change the numerator into
% 600
\begin{equation}
 \frac{1}{d_1 d_2 d_3} = \int^1_0 \ud \alpha_1 \cdots \alpha_3
 \frac{2 \delta \left( 1 - \alpha_1 - \alpha_2 - \alpha_3
  \right)}{\left( \alpha_1 d_1 + \alpha_2 d_2 + \alpha_3 d_3 \right)^3}.
\label{eq.600}
\end{equation}
To obtain the desired form, we expand the denominator in terms of
$\vec k^\perp$ as follows
% 610
\begin{equation}
 D_d = \left(a_d \left( \vec k^\perp \right)^2
 + \vec b^\perp_d \cdot \vec k^\perp + c_d\right)^3,
\label{eq.610}
\end{equation}
which can be rewritten in the form
% 620
\begin{eqnarray}
 D_d & = & \left( a_d \left( \vec k^\perp +
 \frac{\vec b^\perp_d}{2 a_d}\right)^2
 - \frac{(\vec b^\perp_d)^2}{4 a_d} + c_d\right)^3, 
\nonumber \\
 & = & a^3_d \left( {\vec k}'^\perp + \left(M^\perp_d \right)^2\right)^3,
\label{eq.620}
\end{eqnarray}
where for a diagram labeled $d$ we write
% 630
\begin{eqnarray}
 {\vec k}'^\perp & = & \vec k^\perp + \frac{\vec b^\perp_d}{2 a_d},
\nonumber \\
 \left( M_d^\perp \right)^2 & = & -
 \frac{\left(\vec b^\perp_d \right)^2}{4 a^2_d} + \frac{c_d}{a_d}.
\label{eq.630}
\end{eqnarray}
We shift the integration from $\vec k^\perp$ to $\vec k'^\perp$ and
also express the numerator in terms of $\vec k'^\perp$. Then, using
the usual symmetry argument, we see that only even powers of
$\vec k'^\perp$ in the numerator contribute and terms in the
numerator of the form $\left(\vec k'^\perp \cdot \vec a^\perp
\right)\left( \vec k'^\perp \cdot \vec b^\perp \right)$ can be replaced
by  $\frac{1}{2} \, \vec a^\perp \cdot \vec b^\perp \left( \vec
k'^\perp \right)^2$. The final expressions after the shift are rather
complex and thus we will not give them here.

Finally we conclude that we can write $C(d)^{\mu \nu}_{\Gamma
\Gamma}$, $C(d)^\mu_{\Gamma I}$, $C(d)^\nu_{I\Gamma}$ and $C(d)_{II}$
in terms of the following integrals
% 640
\begin{eqnarray}
 \bar{I}^0_3 & = & \int \frac{\ud^2 k^\perp}{(2 \pi)^2}
 \frac{1}{\left( (k^\perp)^2 + (M^\perp)^2 \right)^3}, 
\nonumber \\
 \bar{I}^1_3 & = & \int \frac{\ud^2 k^\perp}{(2 \pi)^2}
 \frac{(k^\perp)^2}{\left((k^\perp)^2 + (M^\perp)^2 \right)^3}, 
\nonumber \\
 \bar{I}^2_3 & = & \int \frac{\ud^2 k^\perp}{(2 \pi)^2}
 \frac{(k^\perp)^4}{\left((k^\perp)^2 + (M^\perp)^2 \right)^3}.
\label{eq.640}
\end{eqnarray}
After the $\vec k^\perp$ integration is done, we also have to integrate
over $k^ +$ and the Feynman parameters.

Since the integral $\bar{I}^2_3$ is divergent, we regulate it. Using DR${}_2$
we write the integral as
% 650
\begin{equation}
 \mu^{2-D}_s \int  \frac{\ud^D k^\perp}{( 2 \pi)^D)} \frac{(k^\perp)^4}{\left(
 (k^\perp)^2 + (M^\perp)^2 \right)^3},
\label{eq.650}
\end{equation}
where $D = 2 - 2 \epsilon$ and a scaling mass $\mu_s$ is introduced
for dimensional reasons. This integral can be done as shown in
Appendix~\ref{sect.B}.  Finally we end up with the following expression
for this integral
% 660
\begin{equation}
 \bar{I}^2_3 = \frac{1}{4 \pi} \left[ \frac{1}{\epsilon} - \gamma -
 \frac{3}{2} + \ln\left(4 \pi\mu^2_s \right) - \ln\left({M^\perp}^2 \right)
 \right].
\label{eq.660}
\end{equation}
The $C(d)^{\mu \nu}_{\Gamma \Gamma}$, $C(d)^\mu_{\Gamma I}$,
$C(d)^\nu_{I \Gamma}$ and $C(d)_{II}$ are linear combinations of these
integrals. When the $\frac{1}{\epsilon} - \gamma - \frac{3}{2} + \ln
\left( 4 \pi \mu^2_s \right)$ terms of the four diagrams are summed,
they cancel out, so we will not take them into account. The
cancellation is proved below.

The divergent terms are the terms in the numerator which are proportional
to $(k'^\perp)^4$. Only the $C(d)^{--}_{\Gamma \Gamma}$ contain such
terms. We will now show that the divergent parts of these terms cancel.
The divergent parts of the four $C(d)^{--}_{\Gamma \Gamma}$'s for
$d=1,\dots,4$ can be written as
% 665
\begin{equation}
 C_d \left[ \frac{1}{\epsilon} -
  \gamma - \frac{3}{2} + \ln \left( 4 \pi \mu^2_s \right) \right],
\label{eq.665}
\end{equation}
with the quantities $C_d$ given by
\begin{eqnarray}
 C_1 & = & \frac{i}{4\pi} \int_0^{{p'}_1^+} \! \frac{\ud k^+}{2 \pi}
 \int_\triangle \! \ud \alpha \, \frac{1}{8 a^3_1}
\nonumber \\
  C_2 & = & - \frac{i}{4\pi} \int_{{p'}_1^+}^{p^+_1} \! \frac{\ud k^+}{2 \pi}
 \int_\triangle \! \ud \alpha \, \frac{1}{8 a^3_2}
\nonumber \\
 C_3 & = & - \frac{i}{4\pi} \int_{{p'}_1^+}^{p^+_1} \! \frac{\ud k^+}{2 \pi}
 \int_\triangle \! \ud \alpha \, \frac{1}{8 a^3_3}
\nonumber \\
 C_4 & = & \frac{i}{4\pi} \int_{p^+_1}^{p^+_1 + p^+_2} \! \frac{\ud k^+}{2 \pi}
 \int_\triangle \! \ud \alpha \, \frac{1}{8 a^3_4},
\label{eq.670}
\end{eqnarray}
where the quantities $a_d$ are defined in Eq.~(\ref{eq.610}).  The
notation $\int_\triangle \! \ud \alpha$ is used for the integration
over the Feynman parameters, which lie in a triangle in $\alpha$-space.

If we substitute the values found for $a_d$, perform the integration
over the Feynman parameters and sum the four integrals, we get
% 680
\begin{eqnarray}
 & &\frac{i}{4 \pi} \frac{1}{8} \left[ \frac{1}{\epsilon} - \gamma - \frac{3}{2}
 + \ln \left( 4 \pi\mu^2_s \right) \right]
 \Bigg[ \int_0^{{p'}_1^+} \! \frac{\ud k^+}{2 \pi} \frac{1}{- p'^+_1
  \left( p^+_1 + p^+_2 \right) p^+_1} 
\nonumber \\
 & & - \int_{{p'}_1^+}^{p^+_1} \! \frac{\ud k^+}{2 \pi} \frac{1}{- p'^+_2
  \left( p^+_1 + p^+_2 \right) p^+_1}
 - \int_{{p'}_1^+}^{p^+_1} \! \frac{\ud k^+}{2 \pi} \frac{1}{- p'^+_2
  \left( p^+_1 - p'^+_1 \right) p^+_1}
\nonumber \\
 & & + \int_{p^+_1}^{p^+_1 + p^+_2} \! \frac{\ud k^+}{2 \pi} \frac{1}{- p'^+_2
  \left( p^+_1 + p^+_2 \right) p^+_2}\Bigg].
\label{eq.680}
\end{eqnarray}
Performing the $k^+$-integration, we find that the
factor multiplying $\left[1/\epsilon - \gamma - 3/2 +
\ln \left( 4 \pi \mu^2 \right) \right]$ vanishes, 
so the divergent terms of the diagrams cancel out.

\subsection{Pauli-Villars regularization}
\label{sect.V.B}

In the previous section, we used DR$_2$ to regulate the divergent
integrals of the four LFTO diagrams. This was only necessary for the
LF calculation, since there are no singularities in the
manifestly covariant case. There are of course other ways to regulate
the divergent terms, for example Pauli-Villars regularization, which is
the topic of this subsection.

In the case of Pauli-Villars regularization, we change the meson
propagator in the following way
% 700
\begin{equation}
 \frac{1}{p^2 - \mu^2 + i \epsilon} \to \frac{1}{p^2 - \mu^2 + i \epsilon}
 - \frac{1}{p^2 - \Lambda^2 + i \epsilon},
\label{eq.700}
\end{equation}
which means we are adding an extra particle to the theory.  All the
integrands are unaffected by this change at low $p$ (since $\Lambda$ is
large), but they are cut off smoothly when $p \gtrsim \Lambda$.

We need only one Pauli-Villars particle because we are only regulating the box
diagram. If we want to regulate more diagrams, we may need more Pauli-Villars 
particles (see for example Brodsky \emph{et al.}~\cite{BHMcC2006}).

\section{Results\label{sect.VI}}
We evaluated the amplitudes using the two regularization schemes given
%in Sect.~\ref{sect.V}. We show that the covariant calculation gives
the in Sect.~\ref{sect.V}. We found that the covariant calculation
gives the same results as the LF calculation, i.e., they are
equivalent.  In the calculations, we have used the following values for
the masses:  $M = 0.94$, $m = 1.44$ and $\mu=0.14$ corresponding to the
masses of the hadrons $N$, Roper resonance, and $\pi$.  These values
satisfy the stability conditions of Sec.  ~\ref{sect.II}.

In all our numerical calculations, we have used Gauss-Legendre
quadratures to perform the integrals over the $\alpha$'s and $k^+$ as well as
an adaptation of Gaussian quadrature for the integrals over ${\vec
k}^\perp$. We have checked the convergence of all numerical calculations.

\subsection{Manifestly covariant calculation}
\label{sect.VI.A}
The matrix elements and the form factors depend on two independent
quantities, namely the Mandelstam variable $s$ (which is directly
related to the total energy) and the scattering angle $\theta$.

\begin{figure}[ptb]
\begin{center}
\includegraphics[width=50mm]{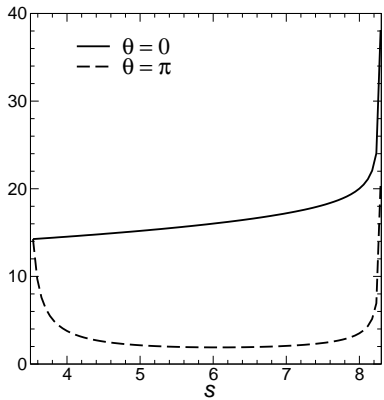}
\end{center}
\caption{The $s$-dependence of $\mathcal{T}_{11}$ in forward and backward 
kinematics in 1+1 dimensions. Covariant calculation. \label{fig.6}}
\end{figure}

\begin{figure}[pbt]
\begin{center}
 \includegraphics[width=50mm]{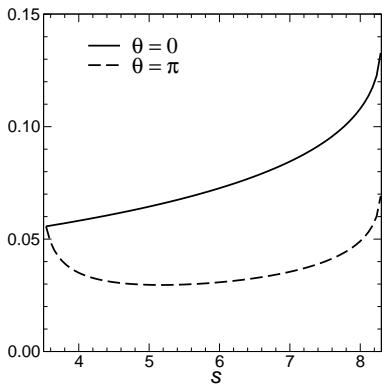}
\end{center}
\caption{The $s$-dependence of $\mathcal{T}_{11}$ in forward and backward 
 kinematics in 3+1 dimensions. Covariant calculation.
\label{fig.7}}
\end{figure}

First we give the results for both kinematics possible for
$\mathcal{T}_{11}$ in 1+1 dimensions. These results are shown in
Fig.~\ref{fig.6}. 
The
lines intersect at the lower threshold $s = 3.53$. This has to be the case,
since at this threshold forward scattering cannot be distinguished from
backward scattering.
For comparison, we give in Fig.~\ref{fig.7} the matrix elements of
$\mathcal{T}_{11}$ in forward and backward scattering for the
3+1-dimensional case.  The behaviour is almost the same as the
1+1-dimensional case, the main difference being the scale.

Of course, we also have in the 3+1 dimensional case the
$\mathcal{T}_{23}$-element in backward scattering, but as its behaviour
is very similar to the behaviour of $\mathcal{T}_{11}$, we do not show
it.

\begin{figure}[tb]
\begin{center}
 $F_1$ \hspace{70mm} $F_2$\\
\vspace{1ex}
\includegraphics[width=60mm]{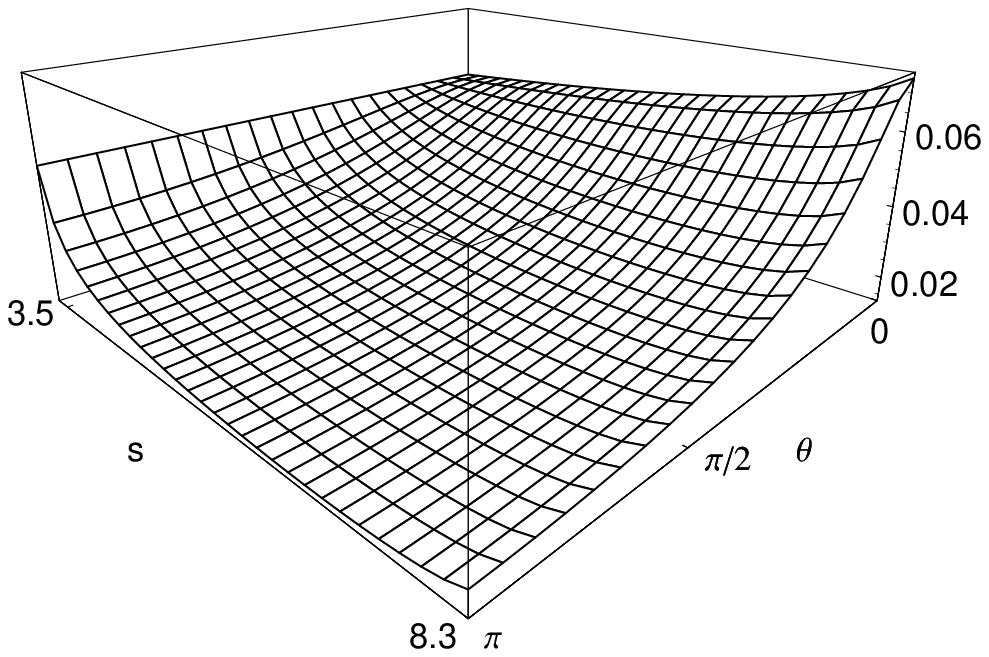} \hspace{10mm}
\includegraphics[width=60mm]{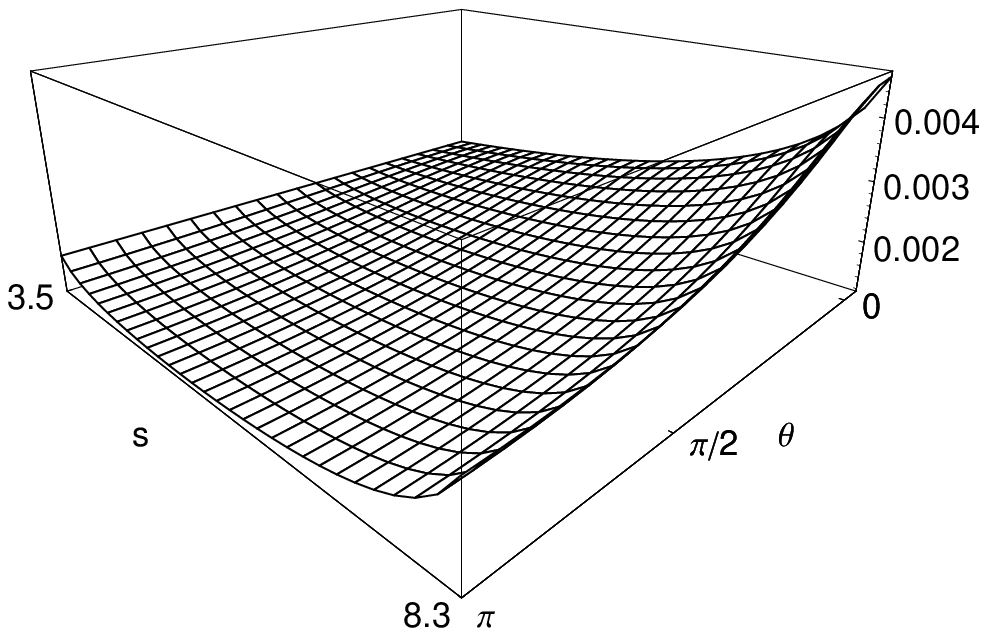}\\
\vspace{1ex}
 $F_3$ \hspace{70mm} $F_4$\\
\vspace{1ex}
\includegraphics[width=60mm]{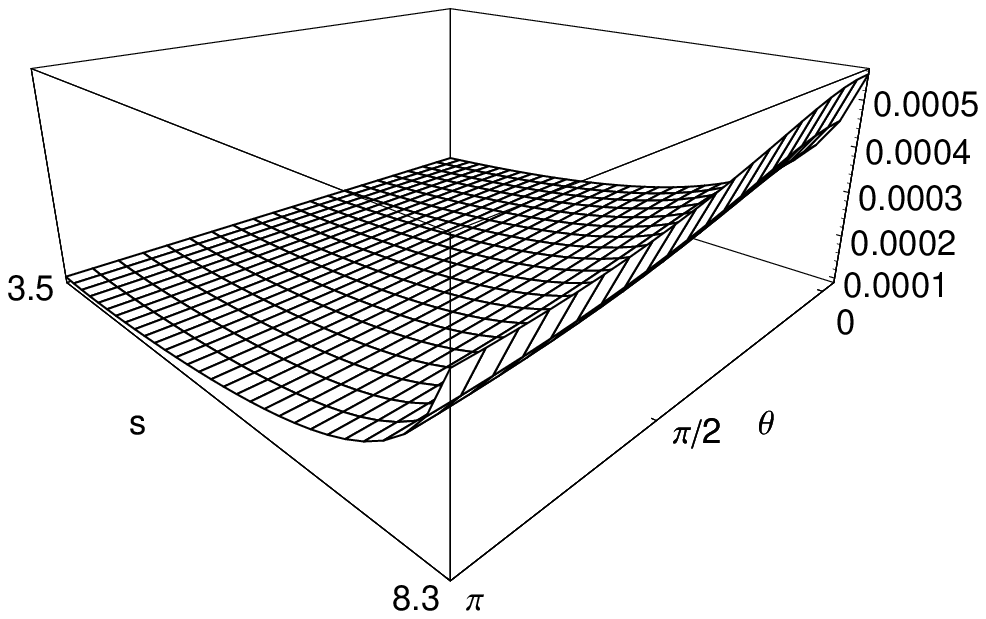} \hspace{10mm}
\includegraphics[width=60mm]{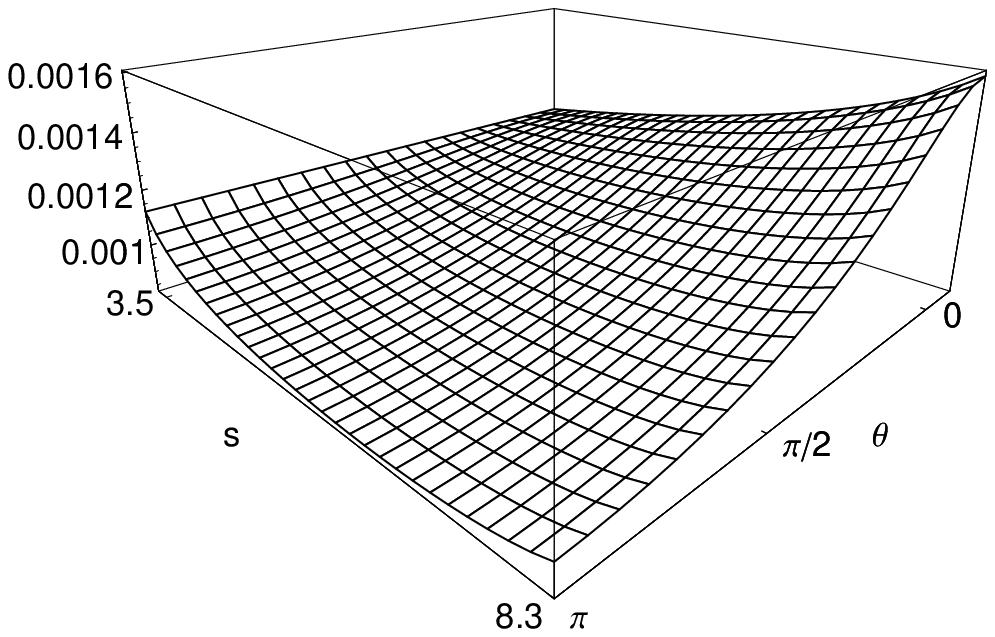}
\end{center}
\caption{The $s$- and $\theta$-dependence of the form factors. Covariant
calculation.  \label{fig.8}}
\end{figure}

In Fig.~\ref{fig.8}, we present our results for the $s$- and
$\theta$-dependence of the four form factors.

\subsection{Light-front calculation using dimensional regularization}
\label{sect.VI.B}

\begin{figure}[ptb]
\begin{center}
 \includegraphics[width=50mm]{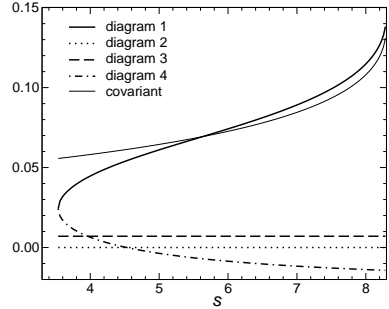}
\end{center}
\caption{The $s$-dependence of the contributions of the four diagrams
to $\mathcal{T}_{11}$ in forward kinematics in 1+1 dimensions.
Light-front calculation. \label{fig.9}}
\end{figure}

\begin{figure}[ptb]
\begin{center}
\vspace{10mm}
 \includegraphics[width=50mm]{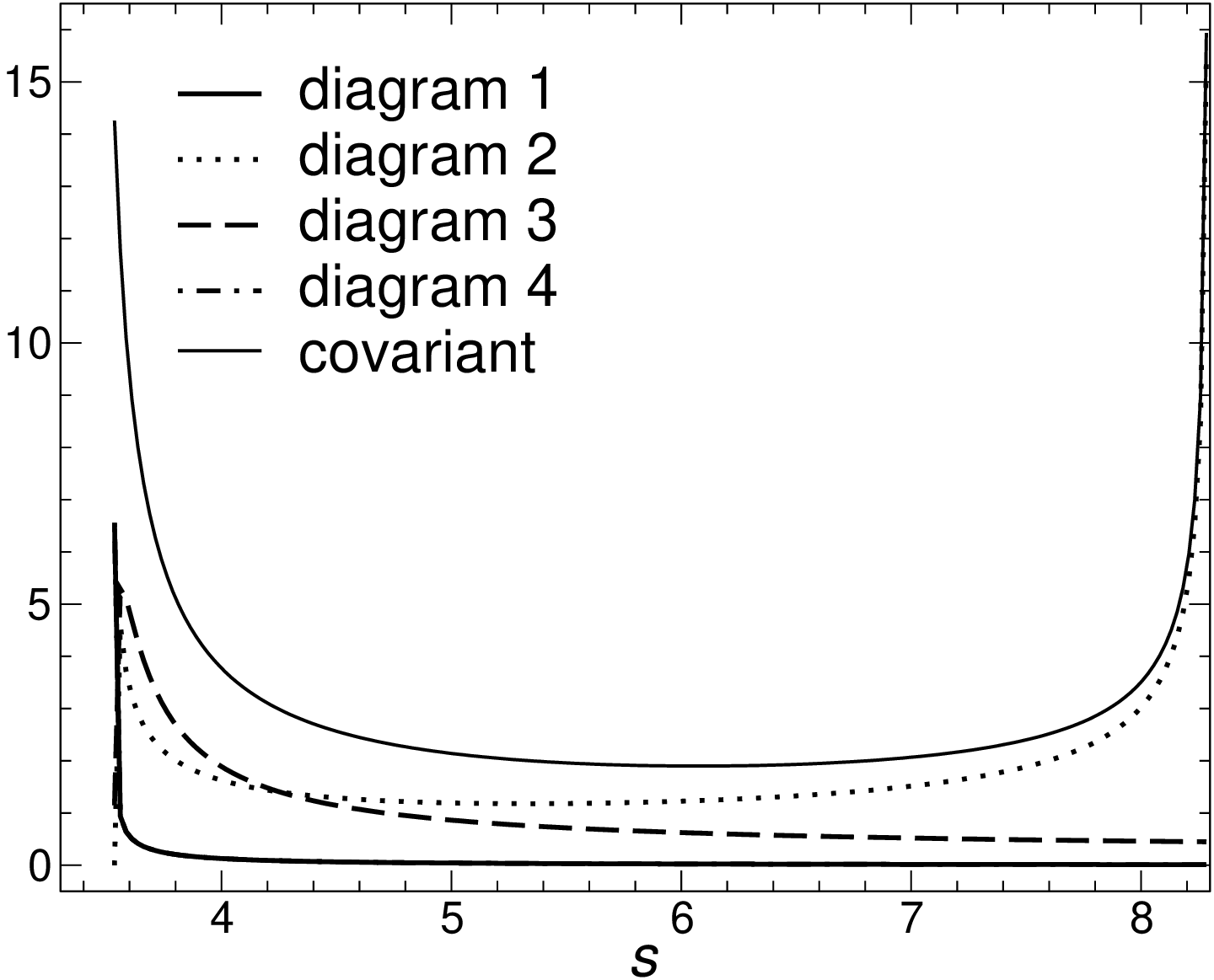}
\end{center}
\caption{The $s$-dependence of the contributions of the four diagrams to 
$\mathcal{T}_{11}$ in backward kinematics in 1+1 dimensions. Light-front
calculation. \label{fig.10}}
\end{figure}

\begin{figure}[ptb]
\begin{center}
 \includegraphics[width=50mm]{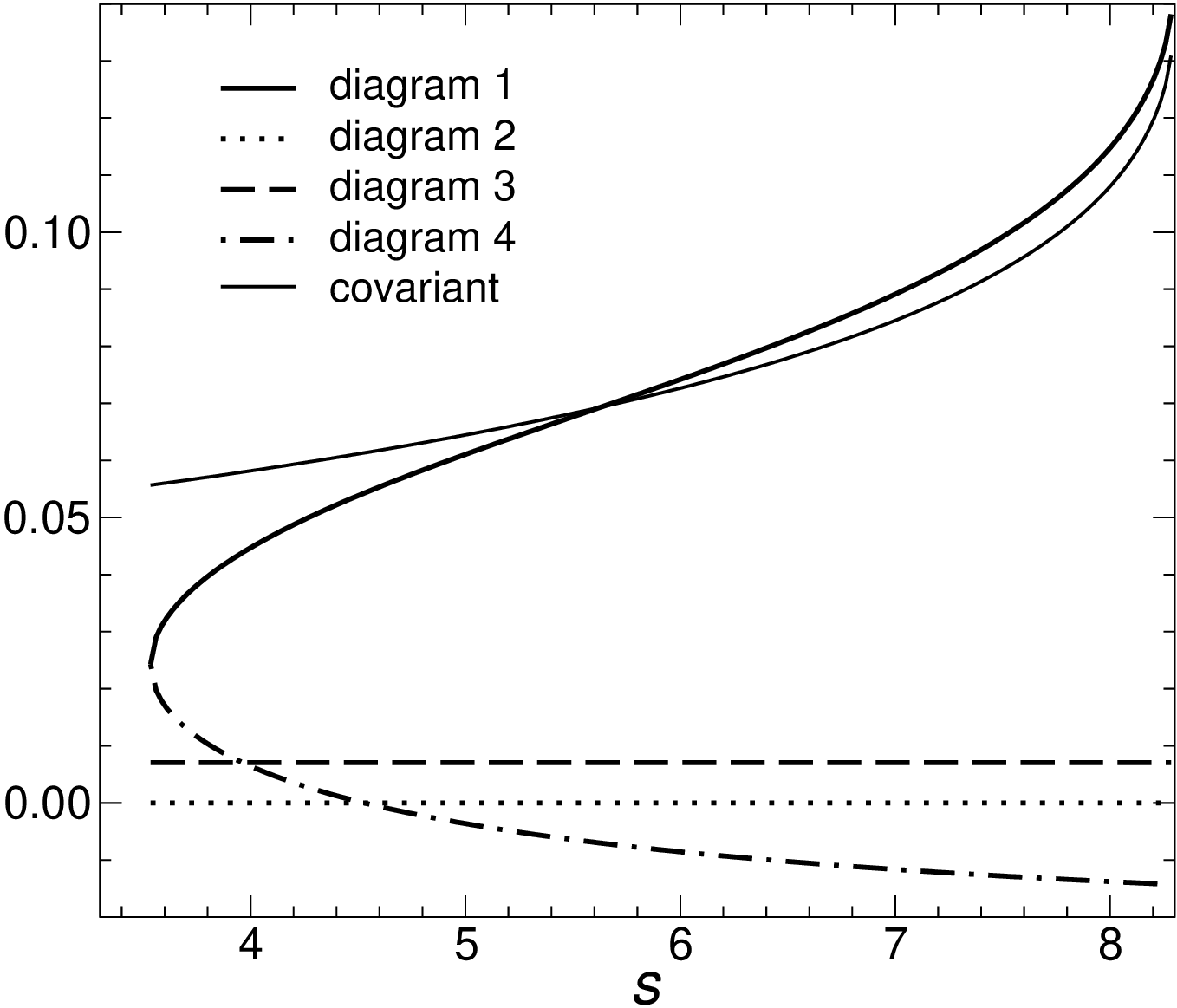}
\end{center}
\caption{The $s$-dependence of the contributions of the four diagrams to 
$\mathcal{T}_{11}$ in forward kinematics in 3+1 dimensions. Light-front
calculation using DR${}_2$. \label{fig.11}}
\end{figure}

\begin{figure}[ptb]
\begin{center}
\includegraphics[width=50mm]{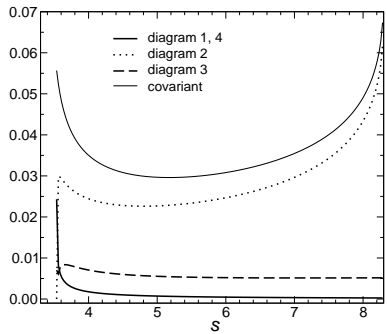}
\end{center}
\caption{The $s$-dependence of the contributions of the four diagrams to 
$\mathcal{T}_{11}$ in backward kinematics in 3+1 dimensions. Light-front
calculation using DR${}_2$.  \label{fig.12}}
\end{figure}

We have calculated the $s$-dependence of the contributions of the four
diagrams to $\mathcal{T}_{11}$ in 1+1 and 3+1 dimensions both in
forward and in backward kinematics. The results are given in
Figs.~\ref{fig.9}~-~\ref{fig.12}. Note that in these cases the
contributions to the form factors cannot be calculated, as the matrix
$O^d_{fi}$ of Sec.~\ref{sect.III.C} is singular.

In Figs.~\ref{fig.9} and \ref{fig.11} the zero mode is clearly visible.
In 1+1 as well as in 3+1-dimensions, diagram 1 dominates at high $s$ in
forward kinematics. In backward kinematics, it is diagram 2 which
%contributes the most. Now we can see that there are more differences
contributes the most. We also see differences
between 1+1 and 3+1 dimensions.  The dominant diagrams are the same,
but otherwise the behaviour of the amplitudes is completely different.

We should note that close to the threshold of the unitarity cut at $s =
8.29$, numerical errors increase and the sum of the LF amplitudes
differs from the covariant amplitude. We have checked that the
covariant amplitudes agree with the LF ones to at least three decimal
places for $s < 5.9$ if we use 20 Gauss points for each dimension in
the numerical integrals. If we increase the number of abscissas to 50,
the accuracy increases to at least 5 decimal places. Thus, we conclude
that the differences are purely due to numerical noise.

\begin{figure}[tb]
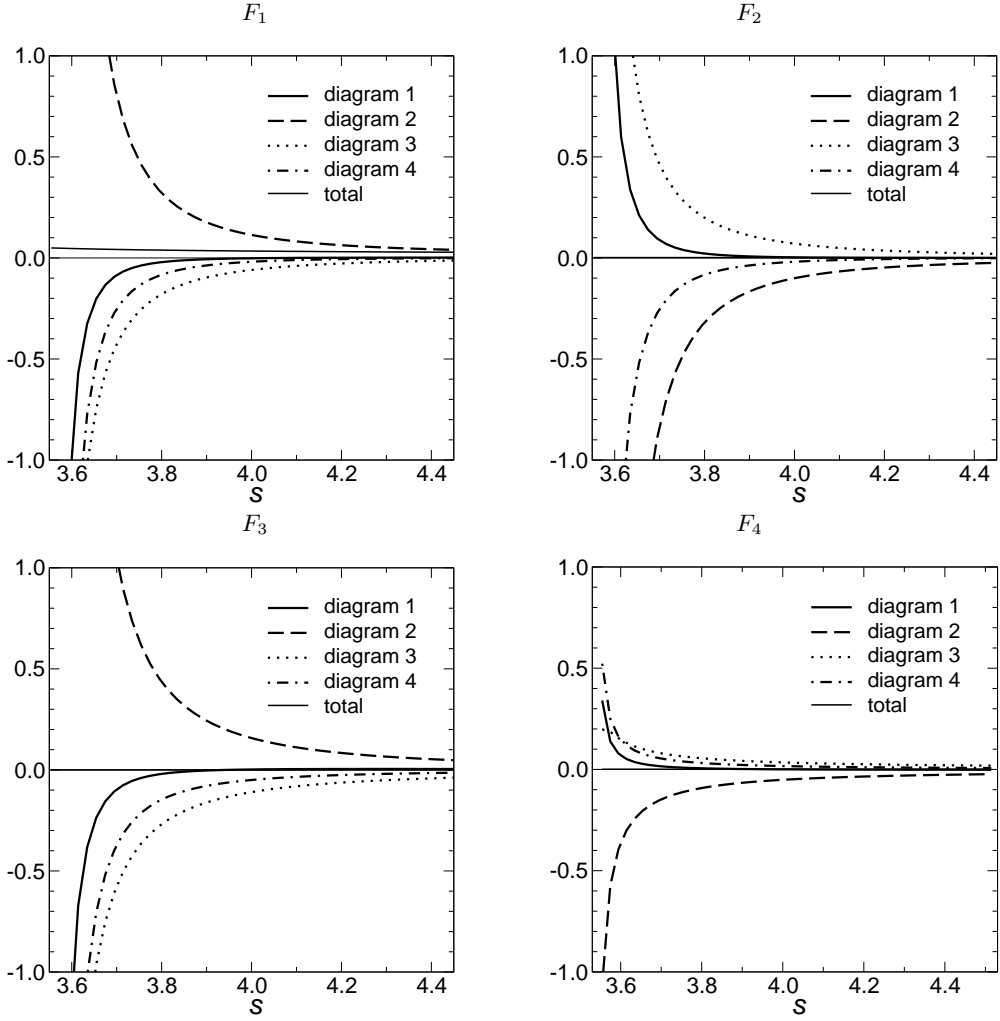

\begin{center}
 $F_1$ \hspace{60mm} $F_2$\\
\vspace{2ex}

\includegraphics[width=60mm]{F1_thetapi2.eps} \hspace{10mm}
\includegraphics[width=60mm]{F2_thetapi2.eps}\\

 $F_3$ \hspace{60mm} $F_4$\\

\vspace{2ex}

\includegraphics[width=60mm]{F3_thetapi2.eps} \hspace{10mm}
\includegraphics[width=60mm]{F4_thetapi2.eps}
\caption{The $s$-dependence of the contributions of the four 
diagrams to the form factors for a value $\theta = \pi/2$. Light-front 
calculation using DR${}_2$. \label{fig.13}}
\end{center}
\end{figure}

We have also calculated the $s$-dependences of the
contributions of the four diagrams to the form factors.  In
Fig.~\ref{fig.13} we show the results for a fixed angle $\theta =
%\pi/2$. We have summed the contributions per diagram to each form
\pi/2$. (We do not show the results for $s> 4.5$, because they show little
structure.) We have summed the contributions per diagram to each form
factor and compared these sums with the covariant form factors.  We
found that the sums and the covariant results are the same, meaning
that the LF calculation is equivalent to the covariant one.

The contributions to the form factors are of the same order of
magnitude as the sum at high $s$ and $\theta$. The LF form factors
diverge for small $s $ and $\theta$, but their sum does not diverge
since the four diagrams compensate each other, rendering the sum
finite. The divergence at small $s$ and $\theta$ of the four diagrams
are to some extent artefacts of the extraction procedure. However, they
become important in a situation where e.g. the stretched box would be
dropped.  They show in a dramatic way the need to include all Fock
sectors that contribute to a certain order in covariant perturbation
theory.

We have obtained our results for the form factors in LFD 
using the extraction procedure in 
Sec.~\ref{sect.III.C}.  In
this extraction procedure, we used the elements of the first row of
$\mathcal{T}$ to obtain our form factors.  Another choice for the four
independent matrix elements of $\mathcal{T}$ gives different results
for the contributions of the four diagrams. The sum, however, does not
change. 
Furthermore, the low $s$- and $\theta$-behaviour will also not
change in the sum, as this behaviour in the individual LFD contribution 
is just an artefact of our extraction procedure.
If one would not take the four-particle intermediate-state into
account, in other words neglect diagram 3, the $s$- and
$\theta$-behaviour would be completely different, most dramatically at small
$s$ and $\theta$.

\subsection{Light-front calculation using Pauli-Villars regularization}

\begin{figure}[ptb]
\begin{center}
 \includegraphics[width=60mm]{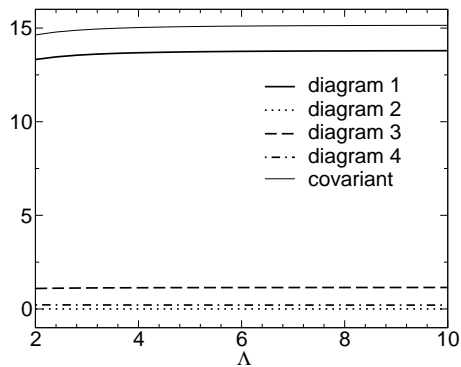}
\end{center}
\caption{The $\Lambda$-dependence of the four contributions to 
$\mathcal{T}_{11}$ in forward kinematics in 1+1 dimensions at $s = 5$.
Light-front calculation with PV subtraction. \label{fig.17}}
\end{figure}

\begin{figure}[ptb]
\begin{center}
 \includegraphics[width=60mm]{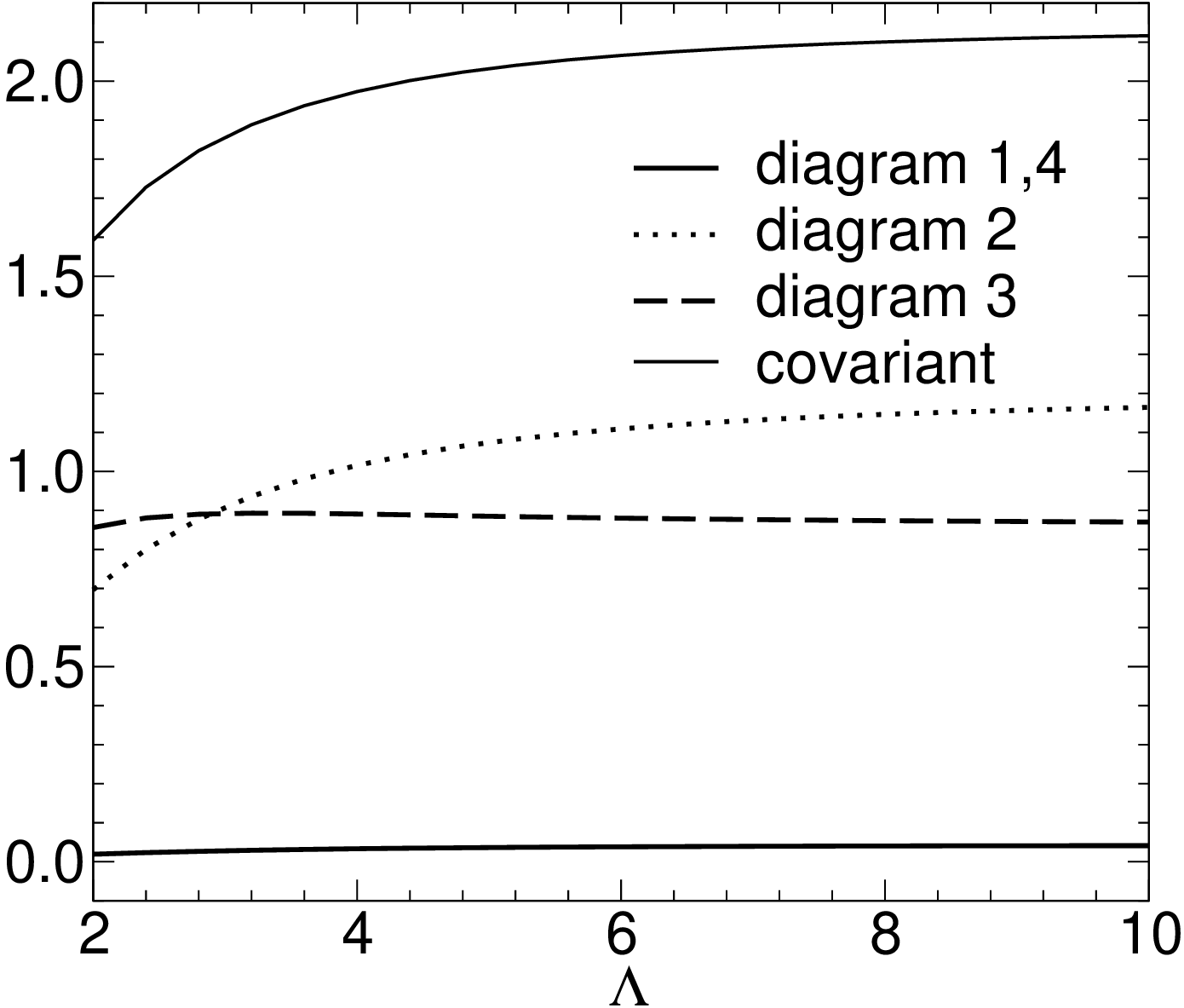}
\end{center}
\caption{The $\Lambda$-dependence of the four contributions to 
$\mathcal{T}_{11}$ in backward kinematics in 1+1 dimensions at $s = 5$.
Light-front calculation with PV subtraction. \label{fig.18}}
\end{figure}

In this subsection, we give our results of the LF calculation using
Pauli-Villars regularization.  In Figs.~\ref{fig.17} and \ref{fig.18}
we present the results in 1+1 dimensions. All the contributions of the
diagrams are convergent, and so is the sum. This has to be the case,
since no regularization is needed in 1+1 dimensions. Furthermore, this
sum is in both cases equal to the covariant result.

\begin{figure}[btb]
\begin{center}
\includegraphics[width=60mm]{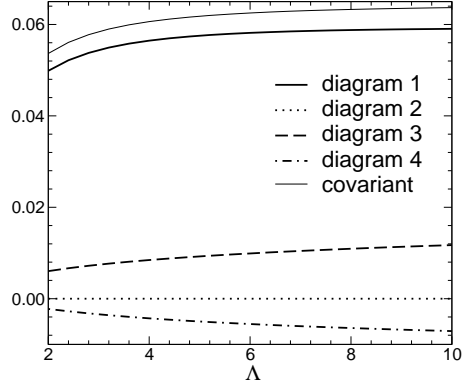}
\end{center}
\caption{The $\Lambda$-dependence of the four contributions to 
$\mathcal{T}_{11}$ in forward kinematics in 3+1 dimensions at $ s = 5$.
Light-front calculation with PV subtraction. \label{fig.19}}
\end{figure}

The results of our calculation of the $\mathcal{T}_{11}$ matrix element
in forward kinematics in 3+1 dimensions at $s = 5$ are shown in
Fig.~\ref{fig.19}.  The logarithmic dependence is clearly visible. 
The divergent parts can be written in the form
\begin{equation}
 C_d \ln \Lambda^2/\mu^2,
\label{eq.715}
\end{equation}
where we use the same coefficients $C_d$, given by Eq.~(\ref{eq.670}),
as in the case of DR${}_2$. Note, however, that this choice is not
unique, as we could also subtract in addition a term of the form $C_d
\times constant$, where $constant$ is independent of the diagram label
$d$. This would not change the sum of the LF amplitudes, because the
sum of the coefficients $C_d$ vanishes. The cancellation of the
logarithmic terms is clearly visible.  Finally, we note that the sum is
equal to the covariant result and that neglecting the stretched box
changes the results considerably, as it is needed to obtain finite
results.

\begin{figure}[p]
\begin{center}
 \includegraphics[width=60mm]{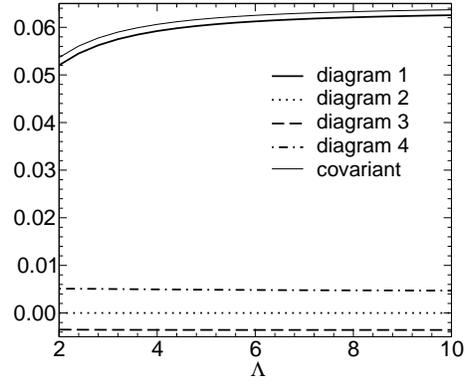}
\end{center}
\caption{The $\Lambda$-dependence of the four contributions to 
$\mathcal{T}_{11}$ in forward kinematics in 3+1 dimensions at $s = 5$. 
In this figure the $\ln \Lambda^2/\mu^2$-term is subtracted.
\label{fig.20}}
\end{figure}

In Fig~\ref{fig.20}, we show the same results, but now we have
subtracted the terms $C_d \ln \Lambda^2 / \mu^2$. Again, one sees that
all diagrams are now convergent so that this divergence is really the
logarithmic divergence found by Van Iersel.

\section{Conclusions\label{sect.VIII}}

We have calculated the box diagram in generalized Yukawa theory, both
in the manifestly covariant formalism and LF quantization. In
the particular case we considered, this diagram can be expressed in
terms of four invariant form factors.

If the calculations are carried out in $1+1$ dimensions, both
approaches lead to finite integrals and the LF amplitudes add
up to the covariant ones. In $3+1$ dimensions, we recovered the
well-known divergences of the LF amplitudes. Using two
completely different regularization methods, we succeeded to remove them
and show that the finite parts of the LF amplitudes again add
up to the covariant ones, while the divergent parts cancel out.  There
is no scheme dependence of the sum of the regularized amplitudes.  One
may change both regularization schemes in such a way that the
individual regulated LF amplitudes in different schemes are
different, but the allowed changes must not change their sum.
There is no anomaly in the box diagram, opposed to the situation
in the triangle diagram of the gauge theory. 

A zero mode occurs, that is revealed only in special kinematical
situations, because it is seen only if there is no plus-momentum
transfer. In the kinematics we adopted, it shows up for special choices
of the scattering angle or the energy, but one might choose the
scattering plane to be perpendicular to the $z$-axis, in which case the
zero mode will occur for all values of the energy and the scattering
angle.

The mechanism for the occurrence of the LF singularities
is easily understood. They are connected to the form of the dispersion
relation for the LF energy. The appearance of the square of
the transverse momentum in the numerator of the LF energy
upsets the usual power counting and opens the possibility for
ulraviolet divergences in LFD that are not present
in the corresponding covariant calculation. From our results, one
understands immediately that logarithmic divergences like the ones
we have encountered in the box diagram will be present in all orders
of perturbation theory.

The taming of the LF singularities can be done only by
including all relevant Fock sectors. In the box diagram, it means that
one may not discard the stretched box. This does not bode well for
%nonperturbative methods that rely on regularization of the kernel,
%e.g., the kernel of the bound-state equation order-by-order.
nonperturbative methods that rely on regularization,
for instance of the kernel of the bound-state equation, order-by-order.
It might turn out that an infinite number of counterterms is needed
to regulate LF Yukawa theory.

For further study, we think that it may be useful to investigate
the double box in order to learn what new singularities may occur.
The double box would correspond to one more iteration of the covariant 
ladder kernel. From this study, we would find out whether the removal
of the singularities of the box diagram is sufficient to regulate
the nonperturbative calculations.

We may also try to attack this problem from a completely different
angle. Namely, we may just do the nonperturbative calculation for
a finite value of the Pauli-Villars mass $\Lambda$ and make
a $\Lambda \to \infty$ extrapolation of the observables, e.g. 
the masses of the bound states. We are convinced that this program would
work if the limit of $\Lambda \to \infty$ and the limit of taking 
an infinite number of diagrams into account commute.

Since the positivity of the LF energy and longitudinal momentum is a
great advantage of the LF formulation, it would be a disappointment if
none of the cures that we would like to try would work.

\appendix

\section{Spinors}
\label{sect.A}

We shall use in this paper helicity spinors. The reason is that in LFD
the LF boosts, which are combinations of pure boosts and rotations,
form a subgroup of the Lorentz group. We get the Kogut and Soper
spinors~\cite{KS1970} by boosting a rest-frame spinor using a LF boost
(see, e.g., Ref.~\cite{BJ2002}). We write them down explicitly for
completeness.  (The symbols $\pm$ refer to $s_z= \pm 1/2$,
respectively.)
% A10
\begin{equation}
 u_{\rm LF}({\bm p};+) = \frac{1}{\sqrt{4\surd 2\,p^+ M}}
 \left(
        \begin{array}{c}
                       \surd 2 \, p^+ + m\\
			p^r \\
                       \surd 2 \, p^+ - m\\
                        p^r
        \end{array}
 \right) ,  \quad
 u_{\rm LF}({\bm p};-) = \frac{1}{\sqrt{4\surd 2\,p^+ M}}
 \left(
        \begin{array}{c}
			-p^l \\
                       \surd 2 \, p^+ + m\\
                        p^l \\
                       -\surd 2 \, p^+ + m
        \end{array}
 \right) .
\label{eq.A10}
\end{equation}

The matrix elements now have the structure 
% A20
\begin{equation}
 I = 
 \left(
        \begin{array}{rl} I_a & I_b  \\ -I^*_b & I^*_a \end{array}
 \right) , \quad
 \Gamma^\mu =
 \left(
 \begin{array}{rl} \Gamma^\mu_a & \Gamma^\mu_b\\
  -{\Gamma^\mu_b}^* & {\Gamma^\mu_a}^* \end{array}
 \right) .
\label{eq.A20}
\end{equation}
The explicit forms of the matrix elements are
% A30
\begin{eqnarray}
 I_a & = & \frac{p^{\prime\, +} + p^+}{2 \sqrt{p^{\prime\, +} p^+}},\quad
 I_b  = 
 \frac{-p^{\prime\, +}p^l + p^+ p^{\prime l}}
 {2M \sqrt{p^{\prime\, +} p^+}}
\nonumber \\
 \Gamma^0_a & = &
 \frac{2 p^{\prime +}p^+ + {\bm p}^{\,\prime}_\perp\cdot {\bm p}_\perp
 + i {\bm p}^{\,\prime}_\perp \times {\bm p}_\perp + M^2}
 {2M \sqrt{2 p^{\prime +} p^+}}, \quad
 \Gamma^0_b  = 
 \frac{p^{\prime l} - p^l}{2 \sqrt{p^{\prime +} p^+}},
\nonumber \\
 \Gamma^x_a & = &
 -\frac{p^{\prime +} p^r + p^+ p^{\prime l}}{2M \sqrt{p^{\prime +} p^+}}, \quad
 \Gamma^x_b  =  -\frac{p^{\prime +} - p^+}{2 \sqrt{p^{\prime +} p^+}},
\nonumber \\
 \Gamma^y_a & = & 
 -i\frac{p^{\prime +} p^r -p^+ p^{\prime l}}{2M \sqrt{p^{\prime +} p^+}}, \quad
 \Gamma^y_b  =  -i \frac{p^{\prime +} - p^+}{2 \sqrt{p^{\prime +} p^+}},
\nonumber \\
 \Gamma^z_a & = &
 \frac{2 p^{\prime +}p^+ - {\bm p}^{\,\prime}_\perp\cdot {\bm p}_\perp
 - i {\bm p}^{\,\prime}_\perp \times {\bm p}_\perp - M^2}
 {2M \sqrt{2 p^{\prime +} p^+}}, \quad
 \Gamma^z_b  = 
 -\frac{p^{\prime l} - p^l}{2 \sqrt{p^{\prime +} p^+}},
\nonumber\\
 \Gamma^+_a & = & \frac{\sqrt{p^{\prime +} p^+}}{M}, \quad \Gamma^+_b = 0,
\nonumber \\
 \Gamma^-_a & = &
 \frac{{\bm p}^{\,\prime}_\perp\cdot {\bm p}_\perp 
 + i {\bm p}^{\,\prime}_\perp \times {\bm p}_\perp + M^2}
 {2M\sqrt{p^{\prime +} p^+}}, \quad
 \Gamma^-_b  = \frac{p^{\prime l} - p^l}{2 \sqrt{p^{\prime +} p^+}}.
\label{eq.A30}
\end{eqnarray}
 
In the kinematics we adopt, we find explicitly for particle $1$
% A40
\begin{eqnarray}
 I_a (1) & = & \frac{2E + p(1+\cos\theta)}{2\sqrt{(E+p)(E+p\cos\theta)}},\quad
 I_b (1) = \frac{(E+p)p \sin\theta}{2M \sqrt{(E+p)(E+p\cos\theta)}};
\nonumber \\
 \Gamma^0_a (1) & = & \frac{(E+p)(E+p\cos\theta)+m^2}
 {2M \sqrt{(E+p)(E+p\cos\theta)}}, \quad
 \Gamma^0_b (1) = \frac{p\sin\theta}{2\sqrt{(E+p)(E+p\cos\theta)}};
\nonumber \\
 \Gamma^x_a (1) & = & \frac{(E+p)p\sin\theta}
 {2M \sqrt{(E+p)(E+p\cos\theta)}}, \quad
 \Gamma^x_b (1) = -\frac{p(1-\cos\theta)}{2\sqrt{(E+p)(E+p\cos\theta)}};
\nonumber \\
 \Gamma^y_a (1) & = & i \frac{(E+p)p\sin\theta}{2M \sqrt{(E+p)(E+p\cos\theta)}}, \quad
 \Gamma^y_b (1) = i\frac{p(1-\cos\theta)}{2\sqrt{(E+p)(E+p\cos\theta)}};
\nonumber \\
 \Gamma^z_a (1) & = & \frac{(E+p)(E+p\cos\theta)-M^2}
 {2M \sqrt{(E+p)(E+p\cos\theta)}}, \quad
 \Gamma^z_b (1) = -\frac{p\sin\theta}{2\sqrt{(E+p)(E+p\cos\theta)}}.
\label{eq.A40}
\end{eqnarray}
For particle $2$ the matrix elements are found by substituting $-p$ for
$p$ in the expressions for particle $1$.

A simple check on these expressions is to go to the forward limit,
$\theta = 0$, where the matrix elements of the identity must reduce to
$\delta_{\lambda'\lambda}$ and of the $\gamma$-matrices to
$p^\mu\,\delta_{\lambda'\lambda}/m$.

\section{Integrals needed in dimensional regularization\label{sect.B}}
We need explicit expressions for the integrals over $\vec k^\perp$
since some of the integrals are divergent. In this appendix, we show
how to regularize these divergences using DR${}_2$.

In LFD, regularization is needed in the transverse directions only. In
dimensional regularization, we calculate the integrals as an analytic
function of the  dimensionality of these directions. We change the
number of dimensions from 2 to D. The integrals are changed according
to
% B10
\begin{equation}
 \int \! \frac{\ud^2 k^\perp}{(2 \pi)^2}
 \frac{\left(\vec k^\perp \right)^{2r}}
 {\left( \left(\vec  k^\perp \right)^{2} + C^2 \right)^s}
 \to
  \mu^{2-D}_s \int \! \frac{\ud^D k^\perp}{(2 \pi)^2}
  \frac{\left(\vec k^\perp \right)^{2r}}
 {\left( \left(\vec  k^\perp \right)^{2} + C^2 \right)^s}
 = \bar I^r_s.
\label{eq.B10}
\end{equation}
Now, we take $D = 2 - 2 \epsilon$ to obtain
% B20
\begin{equation}
 \bar I^r_s = \mu^{2 \epsilon}_s \frac{1}{( 4 \pi)^{D/2}}
 \frac{\Gamma(r+D/2) \Gamma(s -r - D/2)}{\Gamma(D/2) \Gamma(s)} \left(
 \frac{1}{C^2} \right)^{s-r-D/2}.
\label{eq.B20}
\end{equation}
For the divergent integral $\bar I^2_3$, we find
% B30
\begin{eqnarray}
 \bar I^2_3 = \frac{1}{4 \pi} \left[\frac{1}{\epsilon} -  \gamma - \frac{3}{2}
 +  \ln \left( \frac{4 \pi \mu^2_s}{C^2} \right) \right].
\label{eq.B30}
\end{eqnarray}
The other integrals we need are completely regular; 
% B40
\begin{eqnarray}
 \bar I^0_3 & = & \frac{1}{8 \pi} \frac{1}{C^4}, 
\nonumber \\
 \bar I^1_3 & = & \frac{1}{8 \pi} \frac{1}{C^2}.
\label{eq.B40}
\end{eqnarray}


\begin{thebibliography}{99}
%
\bibitem{Dirac} P.A.M. Dirac, 
 Rev. Mod. Phys. {\bf 21}, 392 (1949).
%
\bibitem{BPP} S.J. Brodsky, H.-C. Pauli, and S. Pinsky,
 Phys. Rep. {\bf 301}, 299 (1998).
%
\bibitem{JM} C.-R. Ji and C. Mitchell,
 Phys. Rev. D {\bf 64}, 085013 (2001).
%
\bibitem{JKM} C.-R. Ji, G.-H. Kim, and D.-P. Min,
 Phys. Rev. D {\bf 64}, 025009 (2001);
Phys. Rev. D {\bf 58}, 105020 (1998).
%
\bibitem{BJS} S.J. Brodsky, C.-R. Ji, and M. Sawicki,
 Phys. Rev. D {\bf 32}, 1530 (1985).
%
\bibitem{SFCS2000} J.H.O. Sales, T. Frederico, B.V. Carlson, and P.U. Sauer,
 Phys. Rev. C {\bf 61}, 044003 (2000).
%
\bibitem{SFCS2001} J.H.O. Sales, T. Frederico, B.V. Carlson, and P.U. Sauer,
 Phys. Rev. C {\bf 63}, 064003 (2001).
%
\bibitem{PHW} R.J. Perry, A. Harindranath, and K.G. Wilson,
 Phys. Rev. Lett. {\bf 65}, 2959 (1990).
%
\bibitem{GHPSW} S. G{\l}azek, A. Harindranath, S. Pinsky, J. Shigemitsu, and K.G. Wilson,
 Phys. Rev. D {\bf 47}, 1599 (1993).
%
\bibitem{BDJM} B.L.G. Bakker, M.A. DeWitt, C.-R. Ji, and Y. Mishchenko,
 Phys. Rev. D {\bf 72}, 076005 (2005).
%
\bibitem{BJ2005} B.L.G. Bakker and C.-R. Ji, Phys. Rev. D {\bf 71}, 053005 (2005).
%
\bibitem{MvI2005} M. van Iersel, Few-Body Systems {\bf 36}, 133 (2005).
%
\bibitem{KSW1959} R. Karplus, C.M. Sommerfield, and E.H. Wichmann,
 Phys. Rev. {\bf 114}, 376 (1959)
%
\bibitem{ELOP1966} R.J. Eden, P.V. Landshoff, D.I. Olive, and J.C. Polkinghorne,
 {\em The analytic S-matrix}, (Cambridge UP, Cambridge, 1966)
%
\bibitem{KS1970} J.B. Kogut and D.E. Soper,
 Phys. Rev. D {\bf 1}, 2901 (1970).
%
\bibitem{BJ2002} B.L.G. Bakker and C.-R. Ji,
 Phys. Rev. D {\bf 65}, 073002 (2002).
%
\bibitem{EL2001} E. Leader, {\em Spin in Particle Physics},
(Cambridge University Press, Cambridge, 2001).
%
\bibitem{E1974} K. Erkelenz, Phys. Rept. {\bf 13}, 191 (1974).
%
\bibitem{LB1995} N.E. Ligterink and B.L.G. Bakker, 
Phys. Rev. D {\bf 52}, 5954 (1995).
%
\bibitem{BHMcC2006} S.J. Brodsky, J.R. Hiller, and G. McCartor, 
Ann. Phys. \textbf{321}, 1240 (2006)
%
\bibitem{MvI2004} M. van Iersel,
 \emph{Aspects of Bound-State Calculations in Light-Front Dynamics},
 PhD-thesis, Vrije Universiteit, Amsterdam (2004)

\end{thebibliography}
\end{document}